\documentclass[11pt]{article}
\usepackage{amssymb, amsmath,framed}
\usepackage[colorlinks]{hyperref}
\usepackage[margin=1in]{geometry}
\newcommand{\rem}[1]{}
\newcommand{\de}{{\rm d}}

\newcommand{\bq}{{\mathbf{x}}}

\newcommand{\bK}{{\mathbf{K}}}
\newcommand{\bE}{{\mathbf{E}}}
\newcommand{\bA}{{\mathbf{A}}}
\newcommand{\bB}{{\mathbf{B}}}

\newcommand{\ba}{{\mathbf{a}}}
\newcommand{\balpha}{{\boldsymbol{\alpha}}}
\newcommand{\bu}{{\mathbf{u}}}
\newcommand{\bv}{{\mathbf{v}}}
\newcommand{\bU}{{\boldsymbol{U}}}
\newcommand{\bV}{{\boldsymbol{V}}}
\newcommand{\bW}{{\boldsymbol{W}}}
\newcommand{\bP}{{\mathbf{P}}}
\newcommand{\bp}{\boldsymbol{p}}
\newcommand{\bX}{{\mathbf{X}}}
\newcommand{\bw}{{\mathbf{w}}}
\newcommand{\dvol}{{\de^3\bq\,\de^3\bv}}
\newcommand{\bfi}{\bfseries\itshape}

\newtheorem{theorem}{Theorem}

\newtheorem{corollary}[theorem]{Corollary}
\newtheorem{proposition}[theorem]{Proposition}
\newtheorem{lemma}[theorem]{Lemma}
\newtheorem{remark}{Remark}
\newenvironment{proof}[1][Proof]{\noindent\textbf{#1.} }{\ \rule{0.5em}{0.5em}}

\begin{document}

\title{Euler-Poincar\'e formulation of hybrid plasma models\vspace{-.3cm}}
\author{Darryl D Holm$^1$ and Cesare Tronci$^2$\\
\it\footnotesize $^1$Department of Mathematics, Imperial College London, United Kingdom\\
\it\footnotesize $^2$Section de Math\'ematiques, \'Ecole Polytechnique
F\'ed\'erale de Lausanne, Switzerland\vspace{-.2cm}}
\date{\sf In honor of David Levermore, on the occasion of his 60th birthday.\vspace{-.5cm}}

\maketitle

\begin{abstract}
{\color{black}Three different hybrid Vlasov-fluid systems are derived by applying reduction by symmetry to Hamilton's variational principle. In particular, the discussion focuses on the Euler-Poincar\'e formulation of three major hybrid MHD models, which are compared in the same framework.} These are the current-coupling scheme and two different variants of the pressure-coupling scheme. The Kelvin-Noether theorem is presented explicitly for each scheme, together with the Poincar\'e invariants for its hot particle trajectories. Extensions of Ertel's relation for the potential vorticity and for its gradient are also found in each case, as well as new expressions of cross helicity invariants.
\vspace{-.3cm}
\end{abstract}

\tableofcontents

\section{Introduction}

\subsection{Hybrid Vlasov-fluid models in plasma physics}

Hybrid Vlasov-fluid plasma models contain elements of both
continuum fluids and phase-space probability density. The latter
obeys a Vlasov kinetic equation, which in turn is coupled to the
momentum equation of the background magnetized fluid.

These hybrid kinetic-fluid models arise in several circumstances in modern
plasma physics research, ranging from fusion research
\cite{ParkEtAl} to astrophysical plasmas \cite{WiYiOmKaQu}. 
These circumstances involve the coexistence of a cold fluid
component with an ensemble of energetic particles that require a 
kinetic description. In meeting the challenges presented in such situations, 
the plasma simulation community has begun developing multiscale fluid plasma 
models that allow hybrid descriptions of the two types of flows. This hybrid approach
successfully couples ordinary fluid models for the cold fluid
component to appropriate kinetic equations that govern the phase-space
distribution of the energetic particle species.

One research direction of relevance in applications is the development of hybrid schemes for 
magnetohydrodynamics (MHD) \cite{ParkEtAl}. This development  has split into two promising approaches:
the current-coupling scheme \cite{ParkEtAl,BeDeCh} and
the pressure-coupling scheme \cite{ParkEtAl,FuPark,Cheng,TaBrKi}.
These schemes differ in how the fluid equation is coupled to the kinetic
equation for the hot particles.

Recently, the Hamiltonian formulations of a variety of hybrid
Vlasov-fluid plasma models were developed that led to new theories
of either current-coupled or pressure-coupled hybrid Vlasov-MHD
models  \cite{Tronci2010}. These Hamiltonian formulations
cast considerable light on the energetics of
hybrid Vlasov-fluid plasmas and the relations between their MHD
approximations. In particular,  the current-coupling scheme has been shown to
possess a well defined Hamiltonian structure. However, the pressure-coupling
schemes were found to require additional fluid transport terms in their
accompanying kinetic equations to preserve their corresponding 
Hamiltonian structures, which otherwise would have been lost.

While the Hamiltonian picture of plasma models
provides a powerful tool for energy-conserving properties and may also 
open the way to relevant stability considerations, the question has remained open
whether these models allow a Lagrangian variational formulation
by Hamilton's principle. 
Lagrangian formulations of plasma physics have been successful in several contexts, since they can be easily approached by several approximation methods such as averaging or asymptotic expansions. Well known results of this approach are contained in the Lagrangian wave theory by Dewar \cite{Dewar} and the celebrated {\color{black} guiding center} motion by Littlejohn \cite{Littlejohn}. The key feature of Lagrangian variational formulations is that, in contrast to their Hamiltonian counterparts, the application of essentially any approximation scheme preserves the intrinsic geometrical properties of the resulting dynamics that emerge from its variational structure in the presence of symmetry. 
In particular, the derivations on the Lagrangian side provide a framework in which straight-forward application of asymptotic expansions would still preserve the fundamental circulation and Lie-Poisson properties of these theories. In contrast, asymptotic expansions of the Hamiltonian formulations, for example,  require exceptional care in preserving the Jacobi identity, while applications of asymptotic expansions directly to the equations of motion typically pay no heed to these geometric properties.
An illustrative example of this phenomenon for fluids was given in Camassa, Holm and Levermore \cite{CaHoLe1997} in deriving the ``Lake equations" and {\color{black}``Great Lake equations"}.\footnote{
It was a very good moment in all our careers when Roberto Camassa, Dave Livermore and DDH realized the efficacy of applying asymptotics to Hamilton's principle, while working on the board together one afternoon in Los Alamos.  
}

The above discussion provides a natural motivation for this paper, whose aim  is to present the complementary derivations of the
hybrid fluid models that were introduced in \cite{Tronci2010} from the Hamiltonian side, 
by recovering them on the Lagrangian, or Hamilton's-principle side. 
As in the Hamiltonian Lie-Poisson formulation, the Euler-Poincar\'e approach presented in this paper incorporates the geometric properties that follow from the relabeling symmetry shared by all continuum systems. In addition, the Euler-Poincar\'e variational framework provides a systematic framework for the derivation of other approximate models, which also inherit  these geometric properties from the variational structure.

The new information we gain in this paper in each case is the natural formulation of a Kelvin-Noether circulation theorem and a corresponding Ertel theorem for the potential vorticity. While the comparisons of the Lie-Poisson Hamiltonian properties in paper \cite{Tronci2010} afford insight into the energetics of these theories, the comparisons of their complementary derivations on the Lagrangian side provide distinctions in their circulation laws, and in their Ertel relations for evolution of the potential vorticity and its gradient. Thus, the Euler-Poincar\'e approach affords additional insights into the diagnostics of basic MHD processes in the presence of a hot particle Vlasov component.

Maxwell-Vlasov plasmas have been treated earlier using the
Euler-Poincar\'e variational approach based on applying
symmetry reduction to Hamilton's principle \cite{CeHoHoMa}. The
present work starts with the Low Lagrangian \cite{Low} and systematically
develops a series of approximate Lagrangians for use in
symmetry-reduced Hamilton's principles for re-deriving the hybrid
Vlasov-MHD fluids in \cite{Tronci2010}. These approximate
Lagrangians are shown to admit a variety of symmetry reductions that
produce variants of Kelvin's circulation law for each theory,  together with new expressions for the dynamics of their cross helicities, some of which are found to remain invariant.

 The resulting Euler-Poincar\'e equations recover the equations in \cite{Tronci2010} and illuminate the differences in the interplay between the geometric structure and circulation mechanisms of both the current-coupled and pressure-coupled hybrid Vlasov-MHD models.
The dynamics of the potential vorticity and its gradient are also explained through appropriate generalizations of Ertel's theorem to the hybrid MHD case. These generalizations arise as a direct consequence of the vorticity dynamics produced by the Euler-Poincar\'e equations of motion.

\subsection{Plan of the paper and its main results}

The main content of the paper is, as follows. 
\begin{enumerate}

\item
The remainder of this Introduction reviews the Euler-Poincar\'e construction of the Maxwell-Vlasov equations. Section \ref{kinetic-mf-system} then extends this system to account for the presence of several cold fluid components (kinetic-multifluid system). The reduction process and the resulting circulation laws are presented explicitly, including the Poincar\'e invariant relations that are now obtained from Noether's theorem.

\item
 Section \ref{current-coupling} considers the MHD limit of the kinetic-multifluid system, thereby formulating the Euler-Poincar\'e equations for the hybrid current-coupling MHD scheme. After presenting the Kelvin-Noether theorem, Ertel's relation for the potential vorticity is presented, thereby extending Ertel's theorem for MHD \cite{Hide} to a hybrid model. Also, it is shown that the usual expression of cross helicity is conserved by this hybrid model.

\item Section \ref{pressure-coupling1} presents the Euler-Poincar\'e formulation of the first pressure-coupling hybrid MHD scheme. In this setting, the cold component drives the whole dynamics, so its velocity adds to the mean velocity of the hot particles. This property appears geometrically in the semidirect-product Lie group structure that generates the Euler-Poincar\'e construction. The Kelvin circulation and Ertel potential vorticity relation are derived explicitly, together with a new expression of the cross helicity invariant.

\item Section \ref{pressure-coupling2} focuses on the second pressure-coupling hybrid MHD scheme. In this context, the assumption of a rarefied hot component allows one to neglect the kinetic energy contribution of the corresponding mean flow. Then a decomposition becomes necessary to separate the hot particle velocity from its mean flow. This decomposition produces a \emph{nested} semidirect-product Lie group structure that fits into the Euler-Poincar\'e construction. Explicit expressions for Kelvin circulation, Ertel's theorem and a new cross helicity invariant again result.

\item Finally Section \ref{conclusions} summarizes our main conclusions and discusses the outlook for future research along the present directions.

\end{enumerate}

\subsection{Euler-Poincar\'e formulation of the Maxwell-Vlasov system}

The variational structure of the Maxwell-Vlasov system has been
investigated in many different ways, starting from the pioneering
work by Low \cite{Low}. Since then, several variational formulations
of this system were presented \cite{Pfirsch,PfMo1,PfMo2,YeMo}, which
are mainly based on Eulerian variables. The Low
Lagrangian, however, involves a mixture of Eulerian and Lagrangian
variables. The first variational formulation in terms of purely
Lagrangian variables appeared in \cite{CeHoHoMa}, in which the Low
Lagrangian was modified by the insertion of an extra term. This extra term 
ties the Lagrangian particle velocity to its corresponding
Eulerian coordinate, i.e. $\dot\bq(\bq_0,\bv_0)=\bv(\bq_0,\bv_0)$.
The variational principle was then cast into Euler-Poincar\'e form
\cite{HoMaRa}, by using a reduction process that takes advantage 
of the relevant symmetry properties of the Lagrangian under the Lie group of diffeomorphisms
(smooth invertible maps of both physical space and phase space).

Motivated by the recent results \cite{Tronci2010} on hybrid plasma
models, one may ask whether the variational methods
developed in \cite{CeHoHoMa} would also apply to hybrid Vlasov-fluid
systems. This paper shows that these methods do indeed apply and they provide a systematic
framework in which to develop a fully Lagrangian formulation of the hybrid models.
The resulting theory is again an Euler-Poincar\'e formulation,
which naturally inherits all the Lie-symmetry properties of both
fluid motion and Vlasov kinetic dynamics. These symmetry properties 
then provide the various theorems for circulation and cross-helicity 
that are derived later in the paper. 

The present section introduces the approach that we shall follow throughout
the rest of this paper. In particular, we shall review the
Euler-Poincar\'e variational formulation \cite{CeHoHoMa} of the
Maxwell-Vlasov system
\begin{align}
&\frac{\partial f}{\partial t}+
\bv\cdot\frac{\partial f}{\partial \bq}+{\color{black}\frac{q}{m}}\left(\bE+\bv\times\bB\right)\cdot\frac{\partial f}{\partial \bv}=0
\\
&
\epsilon_0\,\mu_0\,\frac{\partial \bE}{\partial t}=\nabla\times\bB-q\,\mu_0\!\int\!\bv\,f\,\de^3\bv
\,,\hspace{1cm}
\frac{\partial \bB}{\partial t}=-\,\nabla\times\bE
\\
&\epsilon_0\,\nabla\cdot\bE=q\int\!f\,\de^3\bv
\,,\hspace{3.6cm}
\nabla\cdot\bB=0
\end{align}
where $q$ is the particle charge {\color{black} and $m$ its mass}, while $\epsilon_0$ and $\mu_0$ are respectively the dielectric and diamagnetic constants. {\color{black} Moreover, in the standard notation adopted here, $f(\mathbf{x,v},t)$ is the Vlasov distribution on $\Bbb{R}^3\times\Bbb{R}^3$, while $\mathbf{E}(\mathbf{x},t)$ and $\bB(\mathbf{x},t)$ are  the electric field and the magnetic flux, respectively.}

\medskip

\noindent\textbf{Euler-Poincar\'e approach.}\quad The Euler-Poincar\'e approach to the Maxwell-Vlasov system
 is based on an action principle of the type
\[
\delta \int_{t_0}^{t_1} L_{f_0}(\psi,\dot\psi,\Phi,\dot\Phi,\bA,\dot\bA)\,\de t=0
\]
where the Lagrangian $L$ is a functional
\[
L_{f_0}:T\operatorname{Diff}(TM)\times T\mathcal{Q}\to\Bbb{R}
\]
depending on the parameter $f_0\in\operatorname{Den}(TM)$ belonging
to the space of distributions on the tangent bundle $TM$ with local
coordinates $(\bq_0,\bv_0)$.  Here the notation is such that {\color{black} $M$ is the particle configuration space,}
$\psi\in\operatorname{Diff}(TM)$ is an element of the Lie group of
diffeomorphisms of $TM$ and $\mathcal{Q}$ is the space of
electromagnetic potentials $(\Phi,\bA)$, i.e. $\mathcal{Q}=C^\infty(M)\times\Omega ^1(M)$ (where $\Omega ^1(M)$ denotes the space of differential 1-forms on $M$). At this stage, the
variational principle produces Euler-Lagrange equations on
$\operatorname{Diff}(TM)\times \mathcal{Q}$. The explicit form of
the Lagrangian reads as \cite{CeHoHoMa}
\begin{multline}\label{LowLagrangian}
L_{f_0}=\frac12\,m\int \!f_0\Big(
\left|\dot{\bq}(\bq_0,\bv_0)\right|^2+\left|\dot{\bq}(\bq_0,\bv_0)-\bv(\bq_0,\bv_0)\right|^2\Big)\,\de^3\bq_0\,\de^3\bv_0
\\
-q\int \!f_0\Big(
\Phi(\bq(\bq_0,\bv_0))-\dot{\bq}(\bq_0,\bv_0)\cdot\bA(\bq(\bq_0,\bv_0))\Big)
\,\de^3\bq_0\,\de^3\bv_0
\\
+\frac{\epsilon_0}{2}\int
\left|\nabla\Phi+\partial_t\bA\right|^2\de^3\boldsymbol{r}-\frac1{2\mu_0}\int
\left|\nabla\times\bA\right|^2\de^3\boldsymbol{r} \,,
\end{multline}
where {\color{black}the potentials determine the electric field as $\bE=-\nabla\Phi-\partial_t\bA$ and the magnetic flux as $\bB=\nabla\times\bA$. Also,} the map $\psi$ in the action principle above denotes
$\left(\bq(\bq_0,\bv_0),\bv(\bq_0,\bv_0)\right):=\psi(\bq_0,\bv_0)$
and the term
\[
\frac12\,m\int\!
f_0\,\big|\dot{\bq}(\bq_0,\bv_0)-\bv(\bq_0,\bv_0)\big|^2\,\de^3\bq_0\,\de^3\bv_0
\]
allows $\bv(\bq_0,\bv_0)$ to be varied independently and enforces
$\dot\bq=\bv$. Dropping the above term returns precisely the Low
Lagrangian \cite{Low}. 

At this point, the invariance property of the Lagrangian
\eqref{LowLagrangian} is such that
\[
L_{f_0}(\psi,\dot\psi,\Phi,\dot\Phi,\bA,\dot\bA)=L_{f_0\circ\psi^{-1}}(\dot\psi\circ\psi^{-1},\Phi,\dot\Phi,\bA,\dot\bA)=:l(\bX,\Phi,\dot\Phi,\bA,\dot\bA,f)
\]
where we have defined
\begin{equation}\label{X f defs}
\bX:=\dot\psi\circ\psi^{-1}\in\mathfrak{X}(TM)\,,\qquad f:=f_0\circ\psi^{-1}\in\operatorname{Den}(TM)
\end{equation}
and $\mathfrak{X}(TM)$ denotes the Lie algebra of vector fields on $TM$. {\color{black} Notice that the dependance on the identity element $\psi\circ\psi^{-1}$ has been omitted in the reduced Lagrangian.} In this setting, the \emph{reduced} Euler-Poincar\'e Lagrangian $l:\,\mathfrak{X}(TM)\times\operatorname{Den}(TM)\times T\mathcal{Q}\to\Bbb{R}$ produces the equations \cite{CeHoHoMa}
\begin{align}\label{EPMV1}
&\frac{\partial}{\partial t}\frac{\delta l}{\delta
\bX}+\pounds_\bX\,\frac{\delta l}{\delta \bX}=\,
f\,\nabla_{(\bq,\bv)}\frac{\delta l}{\delta f}
\,,\qquad
\frac{\partial f}{\partial t}+\pounds_\bX\, f=\,0
\,,
\\\label{EPMV2}
&
\color{black}
\frac{\partial}{\partial t}\frac{\delta l}{\delta
\dot{\Phi}}-\frac{\delta l}{\delta
\Phi}=0\,,\hspace{3.1cm}
\frac{\partial}{\partial t}\frac{\delta l}{\delta
\dot{\bA}}-\frac{\delta l}{\delta
\bA}=0
\,.
\end{align}
Here, the symbol $\pounds_\bX$ denotes the Lie derivative along the phase-space vector field $\bX\in \mathfrak{X}(TM)$, whose components are given by
\[
\bX(\bq,\bv)=\left(\bu(\bq,\bv),\,\ba(\bq,\bv)\right)
\]
in which
$(\bq,\bv)\in TM$ are the {\color{black} Eulerian}  position-velocity coordinates and
$M=\Bbb{R}^3$, so $\bX\in \mathfrak{X}(\mathbb{R}^6)$. The symmetry-reduced version of the Lagrangian \eqref{LowLagrangian} is
\begin{multline}
l=\int \!f \left(\frac12\,m
\left|\bu\right|^2+\frac12\,m\left|\bu-\bv\right|^2-q\Phi+q\bu\cdot\bA
\right)\dvol
\\
+\frac{\epsilon_0}{2}\int
\left|\nabla\Phi+\partial_t\bA\right|^2\de^3\bq-\frac1{2\mu_0}\int
\left|\nabla\times\bA\right|^2\de^3\bq \,.
\end{multline}
The Maxwell-Vlasov equations are obtained upon applying the variations in the above Lagrangian 
and substituting them into the Euler-Poincar\'e equations \eqref{EPMV1}-\eqref{EPMV2}, as shown in \cite{CeHoHoMa}.

\bigskip

\noindent\textbf{Outlook.}\quad
The remainder of the paper applies the Euler-Poincar\'e approach to the case of
hybrid Vlasov-fluid models that commonly arise in plasma
physics research. After studying a general Vlasov-multifluid system
for the interaction of several fluid plasma components with a hot
particle species, the paper focuses on comparing the Euler-Poincar\'e structures of
current-coupling and pressure-coupling hybrid MHD schemes. In the
latter case, the geometry of the system provides an interesting
example of how the Vlasov distribution function may be transported
by the background fluid, through diffeomorphisms (smooth invertible maps) acting by tangent lifts. 
Our considerations here are restricted to barotropic fluid flows. 

From the strictly mathematical point of view, the case of ideal adiabatic flows that 
transport the specific entropy may be obtained by a straightforward generalization.
However, from the physical viewpoint the role of heat exchange and the effects of an
additional advected quantity should lead to other interesting effects that we intend to discuss elsewhere.
In particular, adiabatic flow effects may be especially interesting for  hybrid fluid {\color{black} drift-kinetic} models, which may be treated in a Lagrangian setting, perhaps by using an approach similar to that for oscillation-center theory, as in \cite{SiKaHo1986}.  For example, this problem might benefit from an exploration of adiabatic invariants that arise from averaging the Hamilton's principle. This is available for Lagrangian theories, but not for Hamiltonian theories, which instead would use Lie series methods. A proof of the equivalence of these theories would also be interesting. This approach follows ideas that go back to Dewar \cite{Dewar}, but now have been further illuminated by the advent of the Euler-Poincar\'e approach to reduction by symmetry for continuum descriptions on the Lagrangian side \cite{HoMaRa}.  

An early step in this direction was already made by Holm, Kupershmidt and Levermore in \cite{HoKuLe1983}, who studied Poisson maps in the Eulerian and Lagrangian descriptions of continuum mechanics. Many of the  concepts from that work, particularly momentum maps from canonical phase spaces to the duals of Lie algebras, remain just as important in the present work as they were then, but here they are applied on the Lagrangian, or Hamilton's principle side for hybrid Vlasov-fluid systems.

\section{Vlasov-multifluid system}\label{kinetic-mf-system}

This section presents the Euler-Poincar\'e formulation of a system composed of several fluid plasma species, {\color{black}each denoted by its label $s=1,\dots,N$,} with an energetic Vlasov component. The Hamiltonian formulation of this system has been presented in \cite{Tronci2010} and its
equations of motion are expressed as
\begin{align}\label{multi-fluid-momentum}
&\rho_s\frac{\partial\bU_s}{\partial
t}+\rho_s\left(\bU_s\cdot\nabla\right)\bU_s =
a_s\rho_s\left(\bE+\bU_s\times\bB\right)-\nabla\mathsf{p}_s
\\
& \frac{\partial\rho_s}{\partial
t}+\nabla\cdot\left(\rho_s\bU_s\right)=0
\\\label{multi-fluid-Vlasov}
& \frac{\partial f}{\partial t}+\bv\cdot\frac{\partial f}{\partial
\bq}+a_h\left(\bE+\bv\times\bB\right)\cdot\frac{\partial f}{\partial
\bv}=0
\\
& \mu_0\epsilon_0\frac{\partial\bE}{\partial
t}=\nabla\times\bB-\mu_0\sum_s
a_s\rho_s\bU_s-\mu_0\,q_h\int\!\bv\,f\,\de^3\bv
\\
&
\frac{\partial\bB}{\partial t}=-\,\nabla\times\bE
\\
& \epsilon_0 \nabla\cdot\bE=\sum_s a_s\rho_s+q_h\int
f\,\de^3\bv \,,\qquad \nabla\cdot\bB=0
\label{multi-fluid-gauge}
\end{align}
In these equations for the Vlasov-multifluid system,  
$a_s=q_s/m_s$ is the charge-to-mass ratio of the fluid species
$s$, while $\rho_s$ and $\bU_s$ are its mass density and velocity, respectively, {\color{black} and $\mathsf{p}_s$ is the scalar partial pressure of species $s$}.
In the above system, the index $h$ denotes the hot
particle component, {\color{black} while each  fluid species $s$ is governed by its own momentum and mass-transport equations. In order to avoid proliferation of indexes, the notation $\nabla$ is relegated to denote only spatial gradients applied to quantities on physical space. Then, gradients on phase space coordinates are denoted by $\nabla_{\!(\bq,\bv)}$, while partial differentiation of phase-space quantities will be denoted by $\partial_\bq$ or $\partial_\bv$.}  

For the case that the fluid component is absent, the
Euler-Poincar\'e formulation of the resulting Maxwell-Vlasov system was presented in
\cite{CeHoHoMa}. On the other hand, an Euler-Poincar\'e formulation of
charged fluids was given in \cite{HoMaRa}. A combination of these two
approaches yields the Euler-Poincar\'e formulation of the
kinetic-multifluid system. Indeed, we shall show that the
equations
\eqref{EPMV1}-\eqref{EPMV2} can be suitably generalized to apply for the Vlasov-multifluid
system. As we shall prove below, this generalization results from the following choice of Lagrangian:
\begin{align}\nonumber
l(\{\bU_s\},\{\rho_s\},\bX,f,\Phi,\dot\Phi,\bA,\dot\bA)=&\frac1{2}\sum_s\int\!
\rho_s\,{\left|\bU_s\right|^2}\,\de^3\bq -\sum_s\int\!\rho_s\big(
\mathcal{U}(\rho_s)+a_s\Phi-a_s\bU_s\cdot\bA\big)\,\de^3\bq
\\\nonumber
&+m_h\int \!f\left(\frac12
\left|\bu\right|^2+\frac12\left|\bu-\bv\right|^2-a_h\Phi+a_h\bu\cdot\bA
\right)\dvol
\\
&+\frac{\epsilon_0}{2}\int
\left|\nabla\Phi+\partial_t\bA\right|^2\de^3\bq-\frac1{2\mu_0}\int
\left|\nabla\times\bA\right|^2\de^3\bq \,,
\label{Kin-mult-Lagrangian}
\end{align}
where $\mathcal{U}(\rho_s)$ denotes the total internal fluid energy, {\color{black} related to partial pressure of species $s$ by $\mathsf{p}_s=\rho_s^2\,\mathcal{U}'(\rho_s)$, while the notation $\{\bU_s\}$ means that that one considers all species $s=1\dots N$ (and analogously for $\{\rho_s\}$)}. In
the special case of a single fluid species, $s=1$ and the above Lagrangian is
defined as a functional
\[
l:\left(\mathfrak{X}(\Bbb{R}^3)\oplus\mathfrak{X}(\Bbb{R}^6)\right)\times
\left(C^\infty(\Bbb{R}^3)^* \times
C^\infty(\Bbb{R}^6)^*\right)\times T\mathcal{Q}(\Bbb{R}^3)\to\Bbb{R}
\]
where $\mathfrak{X}(\Bbb{R}^n)$ denotes the Lie algebra of vector
fields in $\Bbb{R}^n$, the asterisk denotes the distributional dual
space and the tangent space $T\mathcal{Q}(\Bbb{R}^3)$ is constructed
on the space $\mathcal{Q}(\Bbb{R}^3)$ of electromagnetic
4-potentials $(\Phi,\bA)$. In this setting, the advected fluid
quantity is the mass density ${\rho(\bq)\in \operatorname{Den}(\Bbb{R}^3)}$,
while the advected phase-space quantity is the Vlasov distribution
$f(\bq,\bv)\in \operatorname{Den}(\Bbb{R}^6)$.

At this point, in order to use the above Lagrangian, equations
\eqref{EPMV1}-\eqref{EPMV2} must be adapted to the present
case by extending them to account for the presence of the fluid components. In following the treatment in
\cite{HoMaRa,CeHoHoMa}, one writes the Euler-Poincar\'e theorem in
the following general form.\medskip

\begin{theorem}[Euler-Poincar\'e kinetic-multifluid system]
The kinetic-multifluid system
\eqref{multi-fluid-momentum}-\eqref{multi-fluid-gauge} arises from
the Euler-Poincar\'e variational principle
\begin{equation}\label{VP1}
\delta\int_{t_0}^{t_1}\!
\color{black}
l(\{\bU_s\},\{\rho_s\},\bX,f,\Phi,\dot\Phi,\bA,\dot\bA)\,\de t=0
\end{equation}
with the Lagrangian given in \eqref{Kin-mult-Lagrangian} and the variations
\[
\delta\bU_s=\partial_t\mathbf{W}_s-\pounds_{\bU_s}\mathbf{W}_s
\,,\qquad
\delta \bX=\partial_t\mathbf{Z}-\pounds_{\bX}\mathbf{Z}
\,,\qquad
\delta f=-\pounds_{\mathbf{Z}\,}f\,,\qquad\delta\rho_s=-\pounds_{\mathbf{W}_s\,}\rho_s
\]
where $\mathbf{W}_s\in \mathfrak{X}(\Bbb{R}^3)$, $\mathbf{Z}\in \mathfrak{X}(\Bbb{R}^6)$, $\delta\Phi$ and $\delta\bA$ all 
vanish at the endpoints. This variational principle is equivalent to
the Euler-Poincar\'e equations
\begin{align}\label{EP-hybrid1bis}
&\frac{\partial}{\partial t}\frac{\delta l}{\delta
{\bU_s}}+\pounds_{\bU_s}\,\frac{\delta l}{\delta
\bU_s}=\,\rho_s\nabla\frac{\delta l}{\delta \rho_s}
\\\label{EP-hybrid2bis}
&\frac{\partial}{\partial t}\frac{\delta l}{\delta
\bX}+\pounds_\bX\,\frac{\delta l}{\delta \bX}=
f\,\nabla_{(\bq,\bv)}\frac{\delta l}{\delta f}
\\\label{EP-hybrid3bis}
&\frac{\partial \rho_s}{\partial t}+\pounds_{\bU_s}\, \rho_s=\,0\,,
\qquad\frac{\partial f}{\partial t}+\pounds_\bX\, f=0
\\\label{EP-hybrid4bis}
&
\color{black}
\frac{\partial}{\partial t}\frac{\delta l}{\delta
\dot{\Phi}}-\frac{\delta l}{\delta \Phi}=0\,, \qquad
\frac{\partial}{\partial t}\frac{\delta l}{\delta
\dot{\bA}}-\frac{\delta l}{\delta \bA}=0
\,.
\end{align}
\end{theorem}
\begin{proof}
The equivalence between the variational principle \eqref{VP1} and the Euler-Poincar\'e equations \eqref{EP-hybrid1bis}-\eqref{EP-hybrid4bis} follows easily upon repeating the same steps as in \cite{HoMaRa,CeHoHoMa}. In order to derive the equations \eqref{multi-fluid-momentum}-\eqref{multi-fluid-gauge}, one simply computes the functional derivatives of the Lagrangian \eqref{Kin-mult-Lagrangian}. In particular, upon writing $\bX=(\bu,\ba)$, for
the Vlasov kinetic part one has
\[
\frac{\delta l}{\delta \bu}=m_h f\left(2\bu-\bv+a_h\,\bA\right),
\qquad \frac{\delta l}{\delta \ba}=0\,, \qquad \frac{\delta
l}{\delta
f}=\frac{m_h}2|\bu|^2+\frac{m_h}2|\bu-\bv|^2+q_h\bu\cdot\bA-q_h\Phi
\,.
\]
Then, dividing equation \eqref{EP-hybrid2bis} by $f$ yields
\begin{equation}\label{dividebyf}
\frac{\partial}{\partial t}\left(\frac1f\frac{\delta l}{\delta
\bX}\right)+\pounds_\bX\left(\frac1f\frac{\delta l}{\delta
\bX}\right)= \nabla_{(\bq,\bv)}\frac{\delta l}{\delta f} \,.
\end{equation}
Next, projecting onto the second component yields
\[
0=\partial_\bv\bu\cdot\frac1{f}\frac{\delta l}{\delta
\bu}-\partial_\bv\frac{\delta l}{\delta f}=\bu(\bq,\bv)-\bv
\]
so that
\[
\bX(\bq,\bv)=\left(\bv,\ba(\bq,\bv)\right)\,, \qquad \frac{\delta
l}{\delta \bu}= f\left(m_h\bv+q_h\,\bA\right), \qquad
 \frac{\delta
l}{\delta f}=\frac{m_h}2|\bv|^2+q_h\bv\cdot\bA-q_h\Phi .
\]
Upon denoting the particle momentum as $\boldsymbol{p}(\bq,\bv):=m_h\bv+q_h\,\bA(\bq)$ and 
projecting equation \eqref{dividebyf} onto its first component, we obtain
\[
\partial_{t\,}\bp+\left(\bv\cdot\partial_\bq+\ba\cdot\partial_\bv\right)\bp+\partial_\bq\bv\cdot\bp=\partial_\bq\!\left(\frac{m_h}2|\bv|^2+q_h\bv\cdot\bA-q_h\Phi\right)\,,
\]
where $(\bq,\bv)$ are independent coordinates. Standard vector identities then produce the Lorentz force,
\[
\ba(\bq,\bv)=-\,a_h\!\left(\nabla\Phi+\frac{\partial\bA}{\partial
t}\right)+a_h\,\bv\times(\nabla\times\bA)
\,.\]
Therefore, the second equation of \eqref{EP-hybrid3bis} gives the
Vlasov kinetic equation in the form
\[
\frac{\partial f}{\partial t}+\bv\cdot\frac{\partial f}{\partial \bq}
-a_h\left[\left(\nabla\Phi+\frac{\partial\bA}{\partial
t}\right)-\bv\times(\nabla\times\bA)\right]\cdot\frac{\partial f}{\partial \bv} =0
\,.\]

The fluid equations follow easily by inserting the appropriate variational derivatives into the Euler-Poincar\'e equation \eqref{EP-hybrid1bis} and the first of \eqref{EP-hybrid3bis}. For example, one computes
\[
\frac{\delta l}{\delta \bU_s}=\rho_s\bU_s+a_s\,\rho_s\,\bA\,,\qquad
\frac{\delta l}{\delta \rho_s}=-\frac{\de(\rho_s\,\mathcal{U})}{\de \rho_s}-a_s\Phi+a_s\,\bU\cdot\bA
.\]
Next, upon dividing equation \eqref{EP-hybrid1bis} by $\rho$, one obtains
\[
\frac{\partial\bU_s}{\partial t}+\pounds_{\bU_s\,}\bU_s+a_{s\!}\left(\frac{\partial\bA}{\partial t}+\pounds_{\bU_s\,}\bA\right)=-\nabla\mathsf{p}_s-a_s\nabla\Phi+a_s\nabla(\bU_s\cdot\bA)
,\]
where $\mathsf{p}_s=\rho_s^2\,\mathcal{U}'(\rho_s)$ is the scalar partial pressure. Finally, equation \eqref{multi-fluid-momentum} arises from the explicit form of the Lie derivative operation, by using standard vector identities. Analogous arguments also hold for the equations of the electromagnetic potentials.
\end{proof}\\
\begin{remark}[Euler-Poincar\'e reduction] The above theorem follows by an Euler-Poincar\'e reduction process applied to the following unreduced Lagrangian
\begin{align}\nonumber
L_{f_0,\{\rho_{0s}\}}(\{\eta_s\},\{\dot{\eta}_s\},\psi,\dot{\psi},\, \Phi,\dot\Phi,\bA,\dot\bA)=&\frac1{2}\sum_s\int\!
\rho_{0s}(\boldsymbol{a}_0)\,{\left|\dot\eta_s(\boldsymbol{a}_0)\right|^2}\,\de^3\boldsymbol{a}_0
\\\nonumber
&-\sum_s\int\!\rho_{0s}(\boldsymbol{a}_0)\big(
\mathcal{U}(\rho_{0s})+a_s\Phi(\eta_s(\boldsymbol{a}_0))-a_s\,\dot\eta_s(\boldsymbol{a}_0)\cdot\bA(\eta_s(\boldsymbol{a}_0))\big)\,\de^3\boldsymbol{a}_0
\\\nonumber
&+\frac12\,m_h\!\int \!f_0(\bq_0,\bv_0)
\left(
\left|\dot{\boldsymbol{q}}(\bq_0,\bv_0)\right|^2
+
\left|{\dot{\boldsymbol{q}}}(\bq_0,\bv_0)-\boldsymbol{v}(\bq_0,\bv_0)\right|^2
\right)\de^3\bq_0\,\de^3\bv_0
\\\nonumber
&-q_h\int \!f_0(\bq_0,\bv_0)
\Big(
\Phi(\boldsymbol{q}(\bq_0,\bv_0))
+
\dot{\boldsymbol{q}}(\bq_0,\bv_0)\cdot\bA(\boldsymbol{q}(\bq_0,\bv_0))
\Big)\,\de^3\bq_0\,\de^3\bv_0
\\
&+\frac{\epsilon_0}{2}\int
\left|\nabla\Phi+\partial_t\bA\right|^2\de^3\mathbf{r}-\frac1{2\mu_0}\int
\left|\nabla\times\bA\right|^2\de^3\mathbf{r} \,,
\label{Kin-mult-Lagrangian-Unred}
\end{align}
with the notation $(\boldsymbol{q}(\bq_0,\bv_0),\boldsymbol{v}(\bq_0,\bv_0))=\psi(\bq_0,\bv_0)$ for a group action $\psi: \,T\Bbb{R}^3\to T\Bbb{R}^3$. Indeed, the invariance property
\[
L_{f_0,\{\rho_{0s}\}}(\{\eta_s\},\{\dot{\eta}_s\},\psi,\dot{\psi},\, \Phi,\dot\Phi,\bA,\dot\bA)
=
L_{f_0\circ\psi^{-1},\{\rho_{0s}\circ\eta_s^{-1}\}}(\{\dot\eta_s\circ\eta_s^{-1}\},\dot\psi\circ\psi^{-1},\Phi,\dot\Phi,\bA,\dot\bA)
\]
yields the Euler-Poincar\'e Lagrangian
\[
l(\{\bU_s\},\{\rho_s\},\bX,f,\Phi,\dot\Phi,\bA,\dot\bA):=L_{f_0\circ\psi^{-1},\{\rho_{0s}\circ\eta_s^{-1}\}}(\{\dot\eta_s\circ\eta_s^{-1}\},\dot\psi\circ\psi^{-1},\Phi,\dot\Phi,\bA,\dot\bA)
\]
with the notation 
\[
\bU_s=\dot\eta_s\circ\eta_s^{-1}
,\quad
\bX=\dot\psi\circ\psi^{-1}
,\quad\rho_s=\rho_{0s}\circ\eta_s^{-1}
\quad\hbox{and}\quad
f=f_0\circ\psi^{-1}
.\]
This argument follows easily from the treatment in \cite{HoMaRa,CeHoHoMa}. Notice that for the case of a single species $s=1$, the unreduced Lagrangian is of the form
\[
L_{f_0,\,\rho_{0}}:T\operatorname{Diff}(\Bbb{R}^3)\times T\operatorname{Diff}(T\Bbb{R}^3)\times T\mathcal{Q}\to\Bbb{R}\,,
\]
which emphasizes the Lie group structure that underlies the Vlasov-multifluid system \eqref{multi-fluid-momentum}-\eqref{multi-fluid-gauge}.
\end{remark}

\begin{remark}[Kelvin-Noether theorem for the Vlasov multifluid system]
It is easy to verify that equations \eqref{EP-hybrid1bis}-\eqref{EP-hybrid2bis} produce the following circulation conservation laws
\[
\frac{\de}{\de t}\oint_{\gamma_t(\bU_s)}
\big(\bU_s(\bq,t)+a_s\bA(\bq,t)\big)\cdot\de\bq=0\,,\qquad
\frac{\de}{\de t}\oint_{\zeta_t(\bX)}
\big(\bv+a_h\bA(\bq,t)\big)\cdot\de\bq=0\,.
\]
In the first relation the curve $\gamma_t$ moves with the fluid flow, while in the second relation the curve $\zeta_t$ moves with the phase-space vector field $\bX(\bq,\bv)=\big(\bv,\ba(\bq,\bv)\big)$. More explicitly, one can write $\gamma_t=\eta_s(t)\circ\gamma_0$, for a fixed loop $\gamma_0$, and analogously $\zeta_t=\psi(t)\circ\zeta_0$. This difference emphasizes the role of the Poincar\'e invariant associated with the hot particle motion; see \cite{CeHoHoMa}.
\end{remark}

The next sections will consider the Euler-Poincar\'e formulation of hybrid Vlasov-MHD models. In particular, the discussion will focus on two main types of hybrid systems: the current-coupling and pressure-coupling schemes.

\begin{remark}[The Legendre transform] Notice that the Euler-Poincar\'e Lagrangian \eqref{Kin-mult-Lagrangian} is degenerate since $\delta l/\delta\ba=0$. This degeneracy is related to a redundancy in the Euler-Poincar\'e construction, which carries all the information about particle paths that are already encoded in the Vlasov equation. As explained in \cite{CeHoHoMa}, this degeneracy presents some problems when one wants to perform a Legendre transform to obtain the corresponding Hamiltonian description. However, these problems may be overcome by a standard use of Dirac constraints. Similar arguments to those in \cite{CeHoHoMa} also hold for the hybrid models treated in this paper.
\end{remark}

\section{Current-coupling hybrid MHD scheme}\label{current-coupling} 

\subsection{Formulation of the model}
In {\color{black} common} physical situations, one is interested in single-fluid models. In the context of hybrid schemes, it is customary to specialize the system \eqref{multi-fluid-momentum}-\eqref{multi-fluid-gauge} to the two-fluid case and to neglect the inertia of one of the fluid species (electrons). This last approximation is equivalent to taking the limit $m_2\to 0$ for the second species in the total fluid momentum equation. Under this assumption, the sum of the equations \eqref{multi-fluid-momentum} for $s=1,2$ produces
\begin{equation}\label{TotMom}
\rho_1\frac{\partial\bU_1}{\partial
t}+\rho_1\left(\bU_1\cdot\nabla\right)\bU_1 =
\left(a_1\rho_1+a_2\rho_2\right)\bE+\left(a_1\rho_1\bU_1+a_2\rho_2\bU_2\right)\times\bB-\nabla\mathsf{p}_1
\end{equation}
Also, upon assuming neutrality by letting $\epsilon_0\to0$, the electromagnetic fields satisfy the equations
\begin{align}
&\sum_s
a_s\rho_s\bU_s=\frac1{\mu_0}\nabla\times\bB-\,a_h\int\!\bv\,f\,\de^3\bv
\,,\\
&
\frac{\partial\bB}{\partial t}=-\nabla\times\bE
\,,\\
&
\sum_s a_s\rho_s=-q_h\int f\,\de^3\bv
\,,\qquad
\nabla\cdot\bB=0
\,.
\end{align}
Then, equation \eqref{TotMom} becomes
\begin{equation}
\rho\frac{\partial\bU}{\partial
t}+\rho\left(\bU\cdot\nabla\right)\bU = -\left(q_h\int\!
f\,\de^3\bv\right)\bE+\left(\frac1{\mu_0}\nabla\times\bB-\,a_h\int\!\bv\,f\,\de^3\bv\right)\times\bB
- \nabla\mathsf{p}
\,,
\end{equation}
where we have dropped labels for convenience. Finally, inserting
Ohm's ideal law $\bE+\bU\times\bB=0$, the kinetic two-fluid system
becomes
\begin{align}\label{cc-hybrid-momentum}
&\rho\frac{\partial\bU}{\partial
t}+\rho\left(\bU\cdot\nabla\right)\bU = \left(q_h\, \bU\int\!
f\,\de^3\bv-\,q_h\int\!\bv\,f\,\de^3\bv+\frac1{\mu_0}\nabla\times\bB\right)\times\bB
-\nabla\mathsf{p}
\\
& \frac{\partial\rho}{\partial
t}+\nabla\cdot\left(\rho\bU\right)=0
\\
& \frac{\partial f}{\partial t}+\bv\cdot\frac{\partial f}{\partial
\bq}+a_h\left(\bv-\bU\right)\times\bB\cdot\frac{\partial f}{\partial
\bv}=0 \label{cc-hybrid-Vlasov}
\\
& \frac{\partial\bB}{\partial
t}=\nabla\times\left(\bU\times\bB\right) \label{cc-hybrid-end} \,.
\end{align}
This is the same as the current-coupling hybrid scheme presented
in \cite{FuPark,ParkEtAl,BeDeCh}, except that particle
dynamics is governed by the Vlasov equation rather than its
 gyrokinetic counterpart. Notice that the above system does not make
any assumption about the form of the Vlasov distribution for the
energetic particles. Therefore, this system should in principle apply to a
variety of other possible physical situations, as well.

\subsection{Euler-Poincar\'e reduction by symmetry}
We now turn our attention to the Euler-Poincar\'e
formulation of these equations. That is, we ask whether the above current-coupling
system possesses an Euler-Poincar\'e variational principle.
A positive answer is provided by the reduced Lagrangian
\begin{multline}
l(\bU,\rho,\bX,f,\bA)=\frac1{2}\int\!
\rho\,{\left|\bU\right|^2}\,\de^3\bq -\int\!\rho\,
\mathcal{U}(\rho)\,\de^3\bq-\frac1{2\mu_0}\int
\left|\nabla\times\bA\right|^2\de^3\bq
\\
+\int \!f\left(\frac{m_h}2
\left|\bu\right|^2+\frac{m_h}2\left|\bu-\bv\right|^2+q_h\left(\bu-\bU\right)\cdot\bA
\right)\dvol \label{CCS-Lagrangian}
\end{multline}
of the type
\[
l:\big(\mathfrak{X}(\Bbb{R}^3)\oplus\mathfrak{X}(\Bbb{R}^6)\big)\times
C^\infty(\Bbb{R}^3)^*\times\Omega^1(\Bbb{R}^3)\times
C^\infty(\Bbb{R}^6)^*\to\Bbb{R}
\,,
\]
together with the following Euler-Poincar\'e theorem. \medskip
\begin{theorem}
The hybrid current-coupling MHD scheme
\eqref{cc-hybrid-momentum}-\eqref{cc-hybrid-end} arises from the
Euler-Poincar\'e variational principle
\[
\delta\int_{t_0}^{t_1}\! l(\bU,\rho,\bX,f,\bA)\,\de t=0
\]
with the Lagrangian in \eqref{CCS-Lagrangian} and variations given by
\[
\delta\bU=\partial_t\mathbf{W}-\pounds_{\bU}\mathbf{W}
\,,\,\quad
\delta \bX=\partial_t\mathbf{Z}-\pounds_{\bX}\mathbf{Z}
\,,\,\quad
\delta f=-\pounds_{\bX}f\,,\,\quad\delta\rho_s=-\pounds_{\bU}\rho
\,,\,\quad
\delta \bA=-\pounds_{\bU}\bA
\]
where $\mathbf{W}$ and $\mathbf{Z}$ vanish at the endpoints. This
variational principle is equivalent to the Euler-Poincar\'e
equations
\begin{align}\label{EP-CCS1}
&\frac{\partial}{\partial t}\frac{\delta l}{\delta
{\bU}}+\pounds_{\bU}\,\frac{\delta l}{\delta \bU}
=
\,\rho\nabla\frac{\delta l}{\delta \rho}-\frac{\delta l}{\delta
\bA}\times(\nabla\times\bA)
+\left(\nabla\cdot\frac{\delta l}{\delta \bA}\right)\bA
\,,\\
\label{EP-CCS3}
&\frac{\partial \rho}{\partial t}+\pounds_{\bU}\, \rho=\,0\,, \quad\
\frac{\partial \bA}{\partial t}+\pounds_\bU\, \bA=0
\,,\\
\label{EP-CCS2}
&\frac{\partial}{\partial t}\frac{\delta l}{\delta
\bX}+\pounds_\bX\,\frac{\delta l}{\delta \bX}=
f\,\nabla_{(\bq,\bv)}\frac{\delta l}{\delta f}
\,,\\
\label{EP-CCS4}
&\frac{\partial f}{\partial t}+\pounds_\bX\, f=0
\,,
\end{align}
which hold for an arbitrary hybrid Lagrangian.
\end{theorem}
\begin{proof}
The derivation of the Euler-Poincar\'e equations \eqref{EP-CCS1}-\eqref{EP-CCS4} from the Euler-Poincar\'e variational principle can be easily obtained by direct verification \cite{HoMaRa,CeHoHoMa}. In order to derive the current-coupling MHD scheme \eqref{cc-hybrid-momentum}-\eqref{cc-hybrid-end}, one simply computes the functional derivatives and inserts them into the Euler-Poincar\'e equations \eqref{EP-CCS1}-\eqref{EP-CCS4}. In particular, for the Vlasov kinetic part one has
\[
\frac{\delta l}{\delta \bu}=m_h f\left(2\bu-\bv+a_h\,\bA\right),
\qquad \frac{\delta l}{\delta \ba}=0\,, \qquad \frac{\delta
l}{\delta
f}=\frac{m_h}2|\bu|^2+\frac{m_h}2|\bu-\bv|^2+q_h\left(\bu-\bU\right)\cdot\bA
\]
Then, on projecting equation \eqref{EP-hybrid2bis} onto the second
component (recall that $\bX=(\bu,\ba)$), we get
\[
0=\partial_\bv\bu\cdot\frac1f\frac{\delta l}{\delta
\bu}-\partial_\bv\frac{\delta l}{\delta f}=\bu(\bq,\bv)-\bv
\]
so that
\[
\bX(\bq,\bv)=\left(\bv,\ba(\bq,\bv)\right)\,, \qquad \frac{\delta
l}{\delta \bu}=m_h f\left(\bv+a_h\,\bA\right), \qquad
 \frac{\delta
l}{\delta f}=\frac{m_h}2|\bv|^2+q_h\left(\bv-\bU\right)\cdot\bA\,.
\]
Upon denoting $\boldsymbol{p}(\bq,\bv)=m_h\bv+q_h\,\bA(\bq)$ and dividing 
equation \eqref{EP-hybrid2bis} by $f$, one finds
\[
\frac{\partial}{\partial t}\left(\frac1f\frac{\delta l}{\delta
\bX}\right)+\pounds_\bX\left(\frac1f\frac{\delta l}{\delta
\bX}\right)= \nabla_{(\bq,\bv)}\frac{\delta l}{\delta f}
\]
 which when projected onto the first component yields
\[
\partial_{t\,}\bp+\left(\bv\cdot\partial_\bq+\ba\cdot\partial_\bv\right)\bp+\partial_\bq\bv\cdot\bp=\partial_\bq\!\left(\frac{m_h}2|\bv|^2+q_h\left(\bv-\bU\right)\cdot\bA\right)
.
\]
Upon recalling that $(\bq,\bv)$ are independent coordinates and 
using standard vector identities, we can write
\begin{align*}
\ba(\bq,\bv)=&-a_h\!\left(\nabla(\bU\cdot\bA)+\frac{\partial\bA}{\partial
t}\right)+a_h\,\bv\times(\nabla\times\bA)
\\
=&\,a_h\left(\bv-\bU\right)\times(\nabla\times\bA)
\end{align*}
where the bottom line is justified by the second equation in
\eqref{EP-CCS3}. Therefore, equation \eqref{EP-CCS4} returns the
Vlasov kinetic equation \eqref{cc-hybrid-Vlasov} in the form
\[
\frac{\partial f}{\partial t}+\bv\cdot\frac{\partial f}{\partial \bq}
+a_h\big[\!\left(\bv-\bU\right)\times(\nabla\times\bA)\big]\cdot\frac{\partial f}{\partial \bv}=0
\]
with a modified Lorentz force.

We now focus on the fluid part. It suffices to compute
\[
\frac{\delta l}{\delta \bU}=\rho\,\bU-q_h\,n\,\bA \,,\qquad
\frac{\delta l}{\delta
\bA}=-\nabla\times\nabla\times\bA+q_h\left(\bK-n\,\bU\right)
,\qquad\frac{\delta l}{\delta
\rho}=\frac12\left|\bU\right|^2+\rho\,\mathcal{U}^{\,\prime}(\rho)+\mathcal{U}(\rho)
\]
where we have introduced the additional notation
\[
n=\int\!f\,\de^3\bv\,,\qquad\bK=\int\!\bv\,f\,\de^3\bv \,.
\]
At this point, it suffices to insert the above functional
derivatives into equation \eqref{EP-CCS1}, so that
\begin{align}\nonumber
\left(\frac{\partial}{\partial
t}+\pounds_\bU\right)\left(\rho\,\bU-q_h\,n\,\bA\right)=&
\rho\,\nabla\left(\frac12\left|\bU\right|^2+\rho\,\mathcal{U}^{\,\prime}(\rho)+\mathcal{U}(\rho)\right)
+\left(\nabla\times\nabla\times\bA\right)\times\left(\nabla\times\bA\right)
\\
&-q_h\left(\bK-n\,\bU\right)\times\nabla\times\bA
+q_h\,\bA\,\nabla\cdot\left(\bK-n\,\bU\right)
\label{KN1-CCS}
\end{align}
We observe that the zero-th moment of
the Vlasov equation \eqref{cc-hybrid-Vlasov} satisfies $\partial_t
n+\nabla\cdot\bK=0$. Then, making use of the second equation in
\eqref{EP-CCS3} yields
\begin{equation}\label{Aeqn-CCS}
\left(\partial_t+\pounds_\bU\right)(n\,\bA)=-\bA\,\nabla\cdot\left(\bK-n\,\bU\right)
\end{equation}
while expanding the Lie derivatives in \eqref{KN1-CCS} returns the velocity equation
\begin{equation}\label{Ueqn-CCS}
\rho\frac{\partial\bU}{\partial
t}+\rho\left(\bU\cdot\nabla\right)\bU = \left(q_h\,
n\,\bU-\,q_h\bK+\frac1{\mu_0}\nabla\times\bB\right)\times\bB
- \nabla\mathsf{p}
\end{equation}
in which we have substituted $\bB=\nabla\times\bA$ and
{\color{black}$\mathsf{p}=\rho^2\,\mathcal{U}^{\prime}(\rho)$}.
\end{proof}\medskip

\begin{remark}[Euler-Poincar\'e reduction]\label{EPcc-DP}
 Upon following the treatment in \cite{HoMaRa,CeHoHoMa}, one finds that the unreduced Euler-Poincar\'e Lagrangian of the current-coupling scheme is a functional of the type
\[
L_{\rho_0,\bA_0,f_0}:T\operatorname{Diff}(\Bbb{R}^3)\times T\operatorname{Diff}(\Bbb{R}^6)\to\Bbb{R}
\]
where $\times$ denotes direct product. Consequently, 
\[
L_{\rho_0,\bA_0,f_0}=L_{\rho_0,\bA_0,f_0}(\eta,\dot\eta,\psi,\dot\psi)\,.
\]
The reduced Euler-Poincar\'e Lagrangian \eqref{CCS-Lagrangian} is obtained by the reduction process
\[
l(\bu,\bX,\rho,\bA,f)=L_{\rho_0\circ\eta^{-1},\,\bA_0\circ\eta^{-1},\,f_0\circ\psi^{-1}}(\dot\eta\circ\eta^{-1},\dot\psi\circ\psi^{-1})\,.
\]
Here the \emph{advected quantities} $\rho,\bA,f$ are acted on by the corresponding diffeomorphism groups, taking into account their intrisinc tensorial nature, that is $(\rho,\bA,f)\in\operatorname{Den}(\Bbb{R}^3)\times\Omega^1(\Bbb{R}^3)\times\operatorname{Den}(\Bbb{R}^6)$, where $\Omega^1(\Bbb{R}^3)$ denotes the space of differential one-forms on $\Bbb{R}^3$.
\end{remark}

\medskip

\subsection{Discussion}\label{Ertel-CCS}

\textbf{Kelvin circulation law.}\quad Relation \eqref{KN1-CCS} amounts to the following Kelvin circulation law
\begin{multline}
\frac{\de}{\de t}\oint_{\gamma_t(\bU)}\Big(\bU-q_h\frac{n}{\rho}\bA\Big)\cdot\de\bq=q_h\oint_{\gamma_t(\bU)}\frac1\rho\,\Big(\!
\left(\nabla\cdot\left(\bK-n\,\bU\right)\right)\bA
-\left(\bK-n\,\bU\right)\times\bB
\Big)\cdot\de\bq
\\
+
\oint_{\gamma_t(\bU)}\frac1\rho
\left(\nabla\times\bB\right)\times\bB\cdot\de\bq,
\label{KN-CCS}
\end{multline}
which agrees with the corresponding result found in
\cite{Tronci2010}. Notice that the creation of circulation on the
right hand side is generated by the terms involving $\delta l/ \delta\bA$
in the Euler-Poincar\'e equation \eqref{EP-CCS1}. As explained in
\cite{HoMaRa}, these terms comprise a momentum map generated by the action of 
the diffeomorphisms on the cotangent bundle
$T^*\Omega^1(\Bbb{R})$. The presence of these terms is related to
the fact that the non-zero magnetic potential $\bA$ (together with
the mass density $\rho$) \emph{breaks the relabeling symmetry of the
unreduced Lagrangian}, so that
\[
L_{\rho_0,\bA_0,f_0}(\eta,\dot\eta,\psi,\dot\psi)\neq
L_{{\color{black}\rho_0\circ\eta^{-1}},\bA_0,{\color{black}f_0\circ\psi^{-1}}}(\dot\eta\circ\eta^{-1},\psi,\dot\psi)\,.
\]
On the other hand, the Kelvin circulation theorem for the hot
particles reads simply
\[
\frac{\de}{\de t}\oint_{\zeta_t(\bX)}\boldsymbol{p}\cdot\de\bq=0\,,
\]
which recovers the well known preservation of the Poincar\'e
invariant for the hot particle motion.

\bigskip

\noindent
\textbf{Ertel's theorem.}\quad The above Kelvin circulation law identifies the expression of the force 
\begin{equation}\label{KelvinForce-CCS}
\boldsymbol\Psi=\left(\nabla\cdot\left(\bK-n\,\bU\right)\right)\bA
-\left(\bK-n\,\bU\right)\times\bB+\mu_0^{-1}(\nabla\times\bB) \times\bB
\end{equation}
acting on the fluid with momentum $\rho\bU-q_hn\bA$. The above quantity can be used to generalize Ertel's theorem for MHD (see \cite{Hide} and references therein) to the hybrid current-coupling scheme. For simplicity,  consider the incompressible case, so that $\rho\equiv1$ enforces $\nabla\cdot\bU=0$. {\color{black}Next, project the quantity $\rho^{-1}\,n\,\bA$ onto its divergence-free part by defining $\left[\rho^{-1}\,n\,\bA\right]=\rho^{-1}\,n\,\bA+\nabla\varphi$, for a scalar function $\varphi$ such that $\nabla\cdot\left[\rho^{-1}\,n\,\bA\right]=0$. Notice that we keep the density $\rho$ in these relations to provide correct dimensions, while $\rho=1$ for incompressible flows.  Then, upon denoting $D_t=\partial_t+\bU\cdot\nabla$ and $\bar{\boldsymbol{\omega}}=\nabla\times\left(\bU-q_{h\,}{\color{black}\left[\rho^{-1}\,n\,\bA\right]}\right)$} it is easy to see that the curl of equation \eqref{KN1-CCS} produces the generalized Ertel relation
\begin{equation}\label{Ertel}
D_t\left(\bar{\boldsymbol{\omega}}\cdot\nabla\alpha\right)-\left(\bar{\boldsymbol{\omega}}\cdot\nabla\right)D_t\alpha=\nabla\alpha\cdot\nabla\times\boldsymbol\Psi\,,
\end{equation}
where $\alpha$ is an arbitrary smooth function and $\Psi$ is the force expressed by \eqref{KelvinForce-CCS}. The quantity $\bar{\boldsymbol{\omega}}\cdot\nabla\alpha$ is the potential vorticity and the above relation generalizes Ertel's theorem to the current-coupling scheme of hybrid MHD.

\bigskip

\noindent
\textbf{Cross helicities.}\quad 
Upon denoting {$\boldsymbol{\mathcal{V}}=\bU-q_{h\,}\rho^{-1\,}n\bA$}, the following two cross-helicities may now be defined:
\[
\Lambda_1=\int\bU\cdot\bB\,\de^3\bq\,\qquad\Lambda_2=\int\boldsymbol{\mathcal{V}}\cdot\bB\,\de^3\bq
\]
However, while the first is conserved in time, i.e. $\de{\Lambda}_1/\de t=0$, the second satisfies
\[
\frac{\de}{\de t}\Lambda_2=-q_h\frac{\de}{\de t}\int\!\rho^{-1}\, n\,\bA\cdot\bB\,\de^3\bq=q_{h\!}\int\!\rho^{-1}\left(\bA\cdot\bB\right)\nabla\cdot\left(\bK-n\bU\right)\de^3\bq\,,
\]
where the last non-vanishing integral is generated by the term parallel to $\bA$ in equation \eqref{KN1-CCS}.

\section{First pressure-coupling hybrid MHD scheme}\label{pressure-coupling1} 

\subsection{Formulation of the model}

In this section we show how the variational structure of the
previous current-coupling scheme provides a basis for the
Euler-Poincar\'e formulation of a pressure-coupling scheme. This
scheme establishes an equation for the total velocity
\[
\overline{\bU}=\bU+\frac{m_h}\rho\int\bv\,f\,\de^3\bv\,,
\]
under the assumption that the kinetic moment
\begin{equation}\label{AveragedMomentum}
\bK=\int\!\bv\,f\,\de^3\bv
\end{equation}
does not contribute to the total energy
of the system. {\color{black}This assumption can be justified if the energetic component is particularly rarefied, so that its density 
\begin{equation}
n=\int\!f\,\de^3\bv
\end{equation}
is negligible compared to the density $\rho$ of the cold fluid. This is precisely the hypothesis that we shall use in our derivation of the following energy-conserving
pressure-coupling scheme obtained in \cite{Tronci2010}
\begin{align}\label{PCS-MHD1}
&\rho\frac{\partial \bU}{\partial
t}+\rho(\bU\cdot\nabla)\bU=-\nabla{\sf p}-m_h\nabla\cdot\!\int
\!\bv\bv f\,\de^3\bv -\frac1{\mu_0}\bB\times\nabla\times \bB
\\\label{PCS-MHD2}
&\frac{\partial f}{\partial
t}+\left(\boldsymbol{U}+\bv\right)\cdot\frac{\partial f}{\partial
\bq}-\frac{\partial f}{\partial
\bv}\cdot\nabla\boldsymbol{U}\cdot\bv+a_h\,{\bv}\times
\bB\cdot\frac{\partial f}{\partial\bv}=0
\\\label{PCS-MHD3}
&\frac{\partial \rho}{\partial t}+\nabla\cdot(\rho\,\bU)=0
\,,\qquad \frac{\partial \bB}{\partial
t}=\nabla\times\left(\bU\times\bB\right) \,.
\end{align}
Here we have dropped the bar symbol for convenience. Before
proceeding further, we remark that the $\bU$-terms appearing in the
kinetic equation \eqref{PCS-MHD2} differ substantially from the corresponding term in the hybrid MHD model  presented in
\cite{ParkEtAl} (whose Vlasov kinetic equation is replaced by its gyrokinetic approximation). Indeed, the fluid transport term $\bU\cdot\partial_\bq f$ is totally absent in reference \cite{ParkEtAl}, where the circulation force term $\partial_\bv f\cdot\nabla\bU\cdot\bv$ is replaced by the Lorentz force $-q_h\bU\times\bB$ emerging in \eqref{cc-hybrid-Vlasov} as an electric field force (from ideal Ohm's law). More particularly, upon denoting $\Bbb{P}=\int\!\bv\bv\,f\,\de^3\bv$, reference \cite{ParkEtAl} derives a pressure-coupling scheme by 
assuming $\partial_t\bK=-\nabla\cdot\Bbb{P}+a_h(\bK-n\,\bU)\times\bB\simeq0$ in the current-coupling model \eqref{cc-hybrid-momentum}-\eqref{cc-hybrid-end}; then the resulting force balance allows replacing Lorentz forces by the pressure term in the momentum equation \eqref{cc-hybrid-momentum}. These crucial
steps break the energy-conserving nature of the system, as explained in \cite{Tronci2010}. However, notice that the static equilibria of the above equations \eqref{PCS-MHD1}-\eqref{PCS-MHD3} coincide with those of the hybrid model in \cite{ParkEtAl}, provided the hot particles are governed by Vlasov dynamics.}

\subsection{Euler-Poincar\'e reduction by symmetry}
Although the physical approximations leading to the pressure-coupling scheme present some problems that were summarized in \cite{Tronci2010}, we shall see below how the variational approach to the model \eqref{PCS-MHD1}-\eqref{PCS-MHD3} produces an Euler-Poincar\'e system on a semidirect-product Lie group. At the reduced level, we shall prove that the Euler-Poincar\'e Lagrangian is a
functional of the form
\[
l:\big(\mathfrak{X}(\Bbb{R}^3)\,\circledS\,\mathfrak{X}(\Bbb{R}^6)\big)\times
\operatorname{Den}(\Bbb{R}^3)\times\Omega^1(\Bbb{R}^3)\times
\operatorname{Den}(\Bbb{R}^6)\to\Bbb{R}
\]
where the infinitesimal action that is involved in the
semidirect-product Lie algebra $\mathfrak{X}(\Bbb{R}^3)\,\circledS\,\mathfrak{X}(\Bbb{R}^6)$ is given by
\[
\bU\cdot\bX=\pounds_{\bX_\bU}\,\bX\,,\qquad\text{ where
}\qquad\bX_\bU:=\big(\bU,(\bv\cdot\nabla)\bU\big)\qquad\forall
\ \bU\in\mathfrak{X}(\Bbb{R}^3) \,.
\]
This action naturally arises from the tangent-lifted action of
$\operatorname{Diff}(\Bbb{R}^3)$ on $T\Bbb{R}^3=\Bbb{R}^6$, which in
turn generates the natural $\operatorname{Diff}(\Bbb{R}^3)$-action
on $\operatorname{Diff}(\Bbb{R}^6)$ {\color{black}(see Remark \ref{ConjAct} below)}. On the other hand, the \emph{space of the advected quantities} \cite{HoMaRa}
\[
\left(\rho,\bA,f\right)\in
\operatorname{Den}(\Bbb{R}^3)\times\Omega^1(\Bbb{R}^3)\times
\operatorname{Den}(\Bbb{R}^6)
\]
involves the Lie algebra representation
\[
\left(\bU,\bX\right)\cdot\left(\rho,\bA,f\right)=\left(\pounds_\bU\,\rho,\,\pounds_\bU\,\bA,\,\pounds_{\bX+\bX_\bU\,}f\right)
\]
whose associated diamond operation, defined by
\begin{equation}\label{BigDiamond}
\left\langle\left(\frac{\delta l}{\delta \rho},\frac{\delta
l}{\delta \bA},\frac{\delta l}{\delta
f}\right)\diamond\left(\rho,\bA,f\right),\,\left(\bU,\bX\right)\right\rangle:=-\left\langle\left(\frac{\delta
l}{\delta \rho},\frac{\delta l}{\delta \bA},\frac{\delta l}{\delta
f}\right),\Big(\pounds_\bU\,\rho,\,\pounds_\bU\,\bA,\,\pounds_{\bX+\bX_\bU\,}f\Big)\right\rangle,
\end{equation}
will be derived explicitly in what follows.

At this point, the problem has been cast into the standard Euler-Poincar\'e
theory for parameter-dependent Lagrangians $L:TG\times
V^*\to\Bbb{R}$, with the peculiarity that the Lie group
$G$ is a semidirect-product. Indeed, upon replacing $G$ by
$G\,\circledS\,H$, the pressure-coupling scheme will be written as
an Euler-Poincar\'e variational principle on
$T(G\,\circledS\,H)\times V^*$. Upon specializing to the case $G=\operatorname{Diff}(\Bbb{R}^3)$, $H=\operatorname{Diff}(T\Bbb{R}^3)$ and $V^*=\operatorname{Den}(\Bbb{R}^3)\times\Omega^1(\Bbb{R}^3)\times\operatorname{Den}(\Bbb{R}^6)$,  the Euler-Poincar\'e equations
associated to such a Lagrangian can be written as follows on the reduced space $\mathfrak{X}(\Bbb{R}^3)\,\circledS\,\mathfrak{X}({\color{black}\Bbb{R}^6})\times V^*$:
\begin{align}\label{EP-PCS1}
&\frac{\partial}{\partial t}\frac{\delta l}{\delta
{\bU}}+\pounds_{\bU}\,\frac{\delta l}{\delta \bU}=\frac{\delta
l}{\delta \bX}\star\bX+\left(\frac{\delta l}{\delta
\rho},\frac{\delta l}{\delta \bA},\frac{\delta l}{\delta
f}\right)\diamond_1\left(\rho,\bA,f\right)
\\\label{EP-PCS3}
&\frac{\partial \rho}{\partial t}+\pounds_{\bU}\, \rho=\,0\,, \quad\
\frac{\partial \bA}{\partial t}+\pounds_\bU\, \bA=0
\\\label{EP-PCS2}
&\frac{\partial}{\partial t}\frac{\delta l}{\delta
\bX}+\pounds_{\bX+\bX_\bU}\,\frac{\delta l}{\delta \bX}=
f\,\nabla_{(\bq,\bv)}\frac{\delta l}{\delta f}
\\\label{EP-PCS4}
&\frac{\partial f}{\partial t}+\pounds_{\bX+\bX_\bU}\, f=0
\end{align}
where $(\diamond_1)$  in equation (\ref{EP-PCS1}) denotes the $\bU$-component of the diamond
operation {\color{black}defined in \eqref{BigDiamond}}, and the star $(\star)$ operation is defined as
\begin{equation}\label{star-momap}
\left\langle\frac{\delta l}{\delta
\bX}\star\bX,\,\bU\right\rangle:=-\left\langle\frac{\delta l}{\delta
\bX},\,\pounds_{\bX_\bU\,}\bX\right\rangle
.
\end{equation}
Integration by parts yields the more explicit expression,
\begin{align*}
\left\langle\frac{\delta l}{\delta
\bX}\star\bX,\,\bU\right\rangle:=
&
\,-\left\langle\frac{\delta l}{\delta
\bX},\,\pounds_{\bX_\bU\,}\bX\right\rangle
\\
=&\left\langle\pounds_{\bX}\frac{\delta l}{\delta
\bX},\,\big(\bU,(\bv\cdot\nabla)\bU\big)\right\rangle
\\
=& \left\langle\int\!\left(\pounds_{\bX}\frac{\delta l}{\delta
\bX}\right)_{\!1}\!\de^3\bv-\nabla\cdot\!\int\bv\!\left(\pounds_{\bX}\frac{\delta
l}{\delta \bX}\right)_{\!2}\!\de^3\bv,\,\bU\right\rangle,
\end{align*}
so that
\begin{equation}\label{star}
\frac{\delta l}{\delta
\bX}\star\bX=\int\!\left(\pounds_{\bX}\frac{\delta l}{\delta
\bX}\right)_{\!1}\!\de^3\bv-\nabla\cdot\!\int\bv\!\left(\pounds_{\bX}\frac{\delta
l}{\delta \bX}\right)_{\!2}\!\de^3\bv\,,
\end{equation}
where the indices 1 and 2 denote the $\bu$- and the $\ba$-components, respectively.

In order to complete the set of equations
\eqref{EP-PCS1}-\eqref{EP-PCS4}, we shall need a suitable
Euler-Poincar\'e Lagrangian, which is given by
\begin{multline}
l(\bU,\rho,\bX,f,\bA)=\frac1{2}\int\!
\rho\,{\left|\bU\right|^2}\,\de^3\bq -\int\!\rho\,
\mathcal{U}(\rho)\,\de^3\bq-\frac1{2\mu_0}\int
\left|\nabla\times\bA\right|^2\de^3\bq
\\
+\int \!f\left(\frac{m_h}2
\left|\bu\right|^2+\frac{m_h}2\left|\bu-\bv\right|^2+q_h\bu\cdot\bA
\right)\dvol \label{PCS-Lagrangian}
\end{multline}
{\color{black} This Lagrangian is obtained from \eqref{CCS-Lagrangian} by simply neglecting the term $q_h\int\! n\,\bU\cdot\bA\,\de^3\bq$, consistently with the assumption of a rarefied energetic component.}

The simplest starting point involves the kinetic part of the system
\eqref{EP-PCS1}-\eqref{EP-PCS4}, which is composed of the last two
equations, i.e. \eqref{EP-PCS2}-\eqref{EP-PCS4}. Let us start by
calculating the functional derivatives. Upon using similar arguments
as those in the previous section (and especially using the second component
of equation \eqref{EP-PCS2}), in slightly different notation $\bX(\bq,\bv)=\big(\bu(\bq,\bv),\boldsymbol\alpha(\bq,\bv)\big)$ one finds
\[
\bX+\bX_\bU=\big(\bv+\bU,\,\balpha+(\bv\cdot\nabla)\bU\big)\,,
\qquad \frac{\delta l}{\delta \bu}=m_h f\left(\bv+a_h\,\bA\right),
\qquad
 \frac{\delta
l}{\delta f}=\frac{m_h}2|\bv|^2+q_h\bv\cdot\bA\,.
\]
Upon denoting $\boldsymbol{p}(\bq,\bv)=m_h\bv+q_h\,\bA(\bq)$, we divide equation \eqref{EP-hybrid2bis} by $f$ so that
\[
\frac{\partial}{\partial t}\left(\frac1f\frac{\delta l}{\delta
\bX}\right)+\pounds_{\bX+\bX_\bU}\left(\frac1f\frac{\delta l}{\delta
\bX}\right)= \nabla_{(\bq,\bv)}\frac{\delta l}{\delta f}
\]
 and project it onto the first component to obtain
\[
\partial_{t\,}\bp+\big(\left(\bv+\bU\right)\cdot\partial_\bq\big)\bp+(\balpha\cdot\partial_\bv)\bp+\big((\bv\cdot\partial_\bq)\bU\cdot\partial_\bv\big)\bp+\partial_\bq\left(\bv+\bU\right)
\cdot\bp=\partial_\bq\!\left(\frac{m_h}2|\bv|^2+q_h\bv\cdot\bA\right)
.
\]
Then, upon using the second equation in \eqref{EP-PCS3} as well as standard vector identities, one writes
\[
\balpha=a_h\bv\times(\nabla\times\bA)-\nabla\bU\cdot\bv-(\bv\cdot\nabla)\bU
\]
and the vector field $\bX+\bX_\bU$ becomes
\[
\bX+\bX_\bU=\big(\bv+\bU,\,a_h\bv\times(\nabla\times\bA)-\nabla\bU\cdot\bv\big)
\,.
\]
In turn, upon noticing that
$\nabla_{(\bq,\bv)}\cdot\left(\bX+\bX_\bU\right)=0$, this vector field produces the Vlasov kinetic equation
\eqref{EP-PCS2} in the form
\[
\frac{\partial f}{\partial t}+\left(\bv+\bU\right)\cdot\frac{\partial f}{\partial \bq}
-\big(\nabla\bU\cdot\bv-a_h\bv\times(\nabla\times\bA)\big)\cdot\frac{\partial f}{\partial \bv}=0
\]
which is identical to \eqref{PCS-MHD2}.

At this point, one needs to verify that equation \eqref{EP-PCS1} effectively returns the velocity equation \eqref{PCS-MHD1} of the pressure-coupling scheme \eqref{PCS-MHD1}-\eqref{PCS-MHD3}. To this purpose, we shall use the following
\begin{lemma}\label{lemma}
In the special case when
\[
\bX(\bq,\bv)=\big(\bv,\balpha(\bq,\bv)\big)
\qquad \text{ and }\qquad
\frac{\delta l}{\delta
\bX}(\bq,\bv)=\big(\bw(\bq,\bv),0\big)
\]
then
\[
\bX\star\frac{\delta l}{\delta
\bX}=0\,,
\]
for arbitrary vector quantities $\balpha(\bq,\bv)$ and $\bw(\bq,\bv)$.
\end{lemma}
\begin{proof}
The proof follows by direct verification, upon writing the definition of the star operation in (\ref{star-momap})  as follows
\begin{align*}
\left\langle\bX\star\frac{\delta l}{\delta
\bX},\,\bU\right\rangle
:=
\left\langle\frac{\delta l}{\delta
\bX},\Big[\bX_\bU,\,\bX\Big]\right\rangle
=&
\left\langle\frac{\delta l}{\delta
\bX},\Big(\left(\bX_\bU\cdot\nabla_{(\bq,\bv)}\right)\bX-\left(\bX\cdot\nabla_{(\bq,\bv)}\right)\bX_\bU\Big)\right\rangle
\\
=&
\left\langle\bw,\Big(\left(\bX_\bU\cdot\nabla_{(\bq,\bv)}\right)\bv-\left(\bX\cdot\nabla_{(\bq,\bv)}\right)\bU\Big)\right\rangle
\\
=&
\left\langle\bw,\Big(\big((\bv\cdot\nabla)\bU\cdot\partial_{\bv}\big)\bv-\left(\bv\cdot\nabla\right)\bU\Big)\right\rangle
\\
=&
\left\langle\bw,\Big((\bv\cdot\nabla)\bU-\left(\bv\cdot\nabla\right)\bU\Big)\right\rangle=0
\end{align*}
in which the last step uses  integration by parts of the first term.
\end{proof}\bigskip

\noindent
Another result that we shall need is the following formula for the diamond operation in (\ref{star-momap})  
\begin{equation}\label{f-diamond}
\frac{\delta l}{\delta f}\diamond_1
f=\int \!f\,\partial_\bq\frac{\delta l}{\delta
f}\,\de^3\bv-\nabla\cdot \!\int \!f\,\bv\partial_\bv\frac{\delta l}{\delta
f}\,\de^3\bv\,,
\end{equation}
which may be directly verified from its definition
\begin{equation*}
\left\langle \frac{\delta l}{\delta f}\diamond_1
f,\,\bU\right\rangle:=\left\langle
f,\big({\bX_\bU}\cdot\nabla_{(\bq,\bv)}\big)\frac{\delta l}{\delta
f}\right\rangle.
\end{equation*}
 Now, upon recalling the particular form of the variational
derivative 
\[
\frac{\delta l}{\delta f}=\frac12 m_h|\bv|^2+q_h\bv\cdot\bA\,,
\] 
we
calculate
{\color{black}
\begin{align*}
\frac{\delta l}{\delta f}\diamond_1
f=&\
q_h\nabla\bA\cdot\bK-m_h\nabla\cdot\mathbb{P}
-q_h\left(\nabla\cdot\bK\right)\bA-q_h\left(\bK\cdot\nabla\right)\bA
\\
=&\
q_h\bK\times\bB-q_h\left(\nabla\cdot\bK\right)\bA-m_h\nabla\cdot\mathbb{P}\,,
\end{align*}
where we recall the definiton \eqref{AveragedMomentum} of the averaged kinetic momentum and} we have introduced the absolute pressure tensor
\[
\Bbb{P}=\int\!\bv\bv\, f\,\de^3\bv
\,.
\]
Then, upon writing
\[
\frac{\delta l}{\delta
\rho}\diamond_1\rho=\rho\nabla\frac{\delta l}{\delta
\rho}\,,\qquad \frac{\delta l}{\delta
\bA}\diamond_1\bA=-\frac{\delta l}{\delta
\bA}\times\nabla\times\bA+\left(\nabla\cdot\frac{\delta
l}{\delta \bA}\right)\bA
\]
and evaluating
\[
\frac{\delta l}{\delta \bU}=\rho\,\bU \,,\qquad
\frac{\delta l}{\delta
\bA}=-\nabla\times\nabla\times\bA+q_h\bK
,\qquad\frac{\delta l}{\delta
\rho}=\frac12\left|\bU\right|^2+\rho\,\mathcal{U}^{\,\prime}(\rho)+\mathcal{U}(\rho)
\]
we see that equation \eqref{EP-PCS1} returns the velocity equation \eqref{PCS-MHD1} of the pressure-coupling scheme  \eqref{PCS-MHD1}-\eqref{PCS-MHD3}. In conclusion, we have proven the following theorem.
\begin{theorem}
The hybrid pressure-coupling MHD scheme
 \eqref{PCS-MHD1}-\eqref{PCS-MHD3} arises from the
Euler-Poincar\'e variational principle
\[
\delta\int_{t_0}^{t_1}\! l(\bU,\rho,\bX,f,\bA)\,\de t=0
\]
with the reduced Lagrangian
\[
l:\big(\mathfrak{X}(\Bbb{R}^3)\,\circledS\,\mathfrak{X}(\Bbb{R}^6)\big)\times
C^\infty(\Bbb{R}^3)^*\times\Omega^1(\Bbb{R}^3)\times
C^\infty(\Bbb{R}^6)^*\to\Bbb{R}
\]
given in \eqref{PCS-Lagrangian} and variations
\begin{align*}
&\delta\!\left(\bU,\bX\right)=\partial_t\!\left(\mathbf{W},\mathbf{Z}\right)-\left(\pounds_{\bU}\mathbf{W},
\pounds_{\bX_\mathbf{W}}\bX
-
 \pounds_{\bX_\bU}\mathbf{Z}
+
\pounds_{\bX}\mathbf{Z} \right)
\\
& \delta f=-\pounds_{\mathbf{Z}+\bX_\mathbf{W}}f
\,,\,\qquad
\delta\!\left(\rho,
\bA\right)=-\pounds_{\mathbf{W}}\left(\rho,\bA\right)
\end{align*}
where the vector fields $\mathbf{W}\in\mathfrak{X}(\Bbb{R}^3)$ and
$\mathbf{Z}\in\mathfrak{X}(\Bbb{R}^6)$ vanish at the endpoints. This
variational principle is equivalent to the Euler-Poincar\'e
equations \eqref{EP-PCS1}-\eqref{EP-PCS4}, which hold for an
arbitrary hybrid Lagrangian.
\end{theorem}\medskip

\begin{remark}[Conjugation action in semidirect-product Lie groups]\label{ConjAct}
The Lie algebra action that is involved in the semidirect product $\mathfrak{X}(\Bbb{R}^3)\,\circledS\,\mathfrak{X}(\Bbb{R}^6)$ is naturally inherited from the Jacobi-Lie bracket on $\mathfrak{X}(\Bbb{R}^6)$. According to the theory of semidirect-product Lie groups, this action must arise from a group action of $\operatorname{Diff}(\Bbb{R}^3)$ on $\operatorname{Diff}(\Bbb{R}^6)$ that is also a group homomorphism. In other words, $\eta\left(\psi_1\,\psi_2\right)=\eta\left(\psi_1\right)\eta\left(\psi_2\right)$, with $\eta\in\operatorname{Diff}(\Bbb{R}^3)$ and $\psi_1,\psi_2\in\operatorname{Diff}(\Bbb{R}^6)$. In particular, since we can regard $\operatorname{Diff}(\Bbb{R}^3)$ as a subgroup of $\operatorname{Diff}(\Bbb{R}^6)$, one is led to consider the action $\psi\mapsto\eta\circ\psi\circ\eta^{-1}$ which is naturally inherited from the conjugation action in $\operatorname{Diff}(\Bbb{R}^6)$. This action generates the semidirect-product Lie group $\operatorname{Diff}(\Bbb{R}^3)\,\circledS\,\operatorname{Diff}(\Bbb{R}^6)$, whose tangent space at the identity  $\mathfrak{X}(\Bbb{R}^3)\,\circledS\,\mathfrak{X}(\Bbb{R}^6)$ {\color{black} is endowed with}  the Lie bracket 
\[
\left[(\bU,\bX),(\bW,\mathbf{Z})\right]=-(\pounds_{\bU}\mathbf{W},
\pounds_{\bX_\mathbf{W}}\bX
-
\pounds_{\bX_\bU}\mathbf{Z}
+
\pounds_{\bX}\mathbf{Z} )
\,.
\]
More details on semidirect-products of two Lie groups may be found in \cite{MaMiOrPeRa2007,BrGBHoRa2010}.
\end{remark}

\subsection{Discussion}

The Euler-Poincar\'e construction of the first pressure coupling scheme is based on the following proposition:
\begin{proposition}\label{Momap-prop}
The Euler-Poincar\'e equations \eqref{PCS-MHD1}-\eqref{PCS-MHD3} yield
\begin{multline}\label{KM-momap}
\left(\frac{\partial}{\partial t}+\pounds_{\bU}\right)\left(\frac{\delta l}{\delta \bU}-\int\!\frac{\delta l}{\delta \bu}\,\de^3\bv+\int\!\left(\bv\cdot\partial_\bq\right)\frac{\delta l}{\delta \balpha}\,\de^3\bv\right)
=
\rho\,\nabla\frac{\delta l}{\delta
\rho}-\frac{\delta l}{\delta
\bA}\times\nabla\times\bA+\left(\nabla\cdot\frac{\delta
l}{\delta \bA}\right)\bA
\,.
\end{multline}
\end{proposition}
\begin{proof}
The proof is a direct verification, based on relations \eqref{star} and \eqref{f-diamond}. After computing
\begin{align*}
\left(\frac{\partial}{\partial t}+\pounds_{\bU}\right)\left(\frac{\delta l}{\delta \bU}-\int\!\frac{\delta l}{\delta \bu}\,\de^3\bv+\int\right.&\left.\!\!\left(\bv\cdot\partial_\bq\right)\frac{\delta l}{\delta \balpha}\,\de^3\bv\right)
=\frac{\delta l}{\delta
\rho}\diamond_1\rho+\frac{\delta l}{\delta
\bA}\diamond_1\bA
\\
&+\int\!\left(\pounds_{\bX_{\bU}}\frac{\delta l}{\delta \bX}\right)_{\!1}\de^3\bv
-\pounds_{\bU}\int\!\frac{\delta l}{\delta \bu}\,\de^3\bv
\\
&
-\int\!(\bv\cdot\partial_\bq)\left(\pounds_{\bX_{\bU}}\frac{\delta l}{\delta \bX}\right)_{\!2}\de^3\bv
+\pounds_{\bU}\int\!(\bv\cdot\partial_\bq)\frac{\delta l}{\delta \balpha}\,\de^3\bv\,,
\end{align*}
the proof follows immediately from Lemma \ref{lemma-momap} below. 
\end{proof}\medskip

\noindent
\textbf{Kelvin-Noether theorem and its momentum map.}\quad The above relation represents the Lagrangian analogue of an important construction in Lie-Poisson Hamiltonian systems, known as \emph{untangling}. Untangling is accomplished by shifting the momentum by a \emph{momentum map} that takes the Lie-Poisson bracket on the dual of a semidirect-product Lie algebra into the Lie-Poisson bracket dual to a direct-sum Lie algebra. For more details, see Corollary 2.4 in \cite{KrMa1987}. 

It is perhaps not surprising that the very first application of this construction occurred in plasma physics \cite{Holm1,Holm2}. This construction was also used in \cite{Tronci2010}. The momentum map in the present case is the dual $i^*:\mathfrak{X}^*(\Bbb{R}^6)\to\mathfrak{X}^*(\Bbb{R}^3)$ of the Lie algebra inclusion $i:\bU\mapsto\bX_\bU$. 
The result (\ref{KM-momap}) hinges on the following property, which is proven in Appendix \ref{Appendix}: 
\begin{lemma}\label{lemma-momap}
The following map $i^*:\mathfrak{X}^*(\Bbb{R}^6)\to\mathfrak{X}^*(\Bbb{R}^3)$:
\[
i^*\!\left(\frac{\delta l}{\delta \bX}\right)=\int\!\frac{\delta l}{\delta \bu}\,\de^3\bv-\int\!\left(\bv\cdot\partial_\bq\right)\frac{\delta l}{\delta \balpha}\,\de^3\bv\,,
\]
is a momentum map satisfying the relation
\[
i^*\!\left(\pounds_{\bX_\bU}\frac{\delta l}{\delta\bX}\right)=\pounds_{\bU\ }i^*\!\left(\frac{\delta l}{\delta\bX}\right),
\]
for an arbitrary vector field $\bU\in\mathfrak{X}(\Bbb{R}^3)$.
\end{lemma}	
Notice that the momentum map $i^*:\mathfrak{X}^*(\Bbb{R}^6)\to\mathfrak{X}^*(\Bbb{R}^3)$ is different in nature from the star operator ${\star:\mathfrak{X}(\Bbb{R}^6)\times\mathfrak{X}^*(\Bbb{R}^6)\to\mathfrak{X}^*(\Bbb{R}^3)}$ introduced in \eqref{star-momap}. Indeed, while the latter arises from the cotangent lift of the $\operatorname{Diff}(\Bbb{R}^3)$-action on $\mathfrak{X}(\Bbb{R}^6)$, the momentum map $i^*$ arises from the $\operatorname{Diff}(\Bbb{R}^3)$-action on $\operatorname{Diff}(\Bbb{R}^6)$, which is given by conjugation, as explained in Remark \ref{ConjAct}. The momentum map property of $i^*$ can be easily verified since the inclusion $i:\mathfrak{X}(\Bbb{R}^3)\hookrightarrow\mathfrak{X}(\Bbb{R}^6)$ is the dual of a Lie algebra homomorphism, i.e. 
\begin{align*}
\left[\bX_\bU,\bX_\bW\right]=\left[i(\bU),i(\bW)\right]=i\!\left(\left[\bU,\bW\right]\right)=\bX_{\left[\bU,\bW\right]}\,,
\end{align*}
where $\left[\cdot,\cdot\right]$ denotes minus the Jacobi-Lie bracket on $\mathfrak{X}(\Bbb{R}^3)$ or $\mathfrak{X}(\Bbb{R}^6)$, depending on the context. The proof in Appendix \ref{Appendix} shows explicitly that $i^*$ satisfies the definition of momentum map.

An immediate consequence of Proposition \ref{Momap-prop} is the following circulation law for the hybrid scheme \eqref{PCS-MHD1}-\eqref{PCS-MHD3}, which recovers the previous results in \cite{Tronci2010}.\medskip

\begin{corollary}[Kelvin circulation law]\label{KN-PCS1} The pressure-coupling MHD scheme
\eqref{PCS-MHD1}-\eqref{PCS-MHD3} possesses the following equivalent
circulation theorems
\begin{align}\label{KN1}
&\frac{\de}{\de
t}\oint_{\gamma_t(\bU)}\bU\cdot\de\bq=-\oint_{\gamma_t(\bU)}\frac1\rho
\left(\frac1{\mu_0}\bB\times\nabla\times
\bB+m_h\nabla\cdot\!\int \!\bv\bv f\,\de^3\bv\right)\cdot\de\bq
\\\label{KN2}
&\frac{\de}{\de t}\oint_{\gamma_t(\bU)}\left(\bU-\frac1\rho\int
f\,\bp\,\de^3\bv\right)\cdot\de\bq=-\oint_{\gamma_t(\bU)}\frac1\rho
\Big(\bB\times\left(\mu_0^{-1}\nabla\times
\bB-q_h\bK\right)-q_h(\nabla\cdot\bK)\bA\Big)\cdot\de\bq \,.
\end{align}
\end{corollary}
\begin{proof}
Upon considering the Euler-Poincar\'e Lagrangian \eqref{PCS-Lagrangian}, relation \eqref{KN1} is implied by the Euler-Poincar\'e theorem. See \cite{HoMaRa}) and the equation of motion \eqref{EP-PCS1}. On the other hand, relation \eqref{KN2} is an immediate consequence of equation \eqref{KM-momap} in Proposition \ref{Momap-prop}. 
\end{proof}

\noindent
{\color{black}Notice that taking the difference of the above circulation laws yields
\[
\frac{\de}{\de t}\oint_{\gamma_t(\bU)}\!\left(\frac1\rho\int\!
f\,\bp\ \de^3\bv\right)\cdot\de\bq=-\oint_{\gamma_t(\bU)}\frac1\rho
\Big(m_h\nabla\cdot\!\int \!\bv\bv f\,\de^3\bv-q_h\bK\times\bB+q_h(\nabla\cdot\bK)\bA\Big)\cdot\de\bq
\]
where we recall the relation $\boldsymbol{p}=m_h\bv+q_h\,\bA$. Thus, upon considering \eqref{EP-PCS2} and the zero-th moment equation $\partial_t n+\nabla\cdot(n\bU)=-\nabla\cdot\bK$ associated to \eqref{PCS-MHD2}, we have
\[
\frac{\de}{\de t}\oint_{\gamma_t(\bU)}\frac\bK\rho\cdot\de\bq=-\oint_{\gamma_t(\bU)}\frac1\rho
\Big(\nabla\cdot\!\int \!\bv\bv f\,\de^3\bv-a_h\bK\times\bB\Big)\cdot\de\bq
\,.
\]
Moreover,} the above fluid circulation laws are accompanied by preservation of the Poincar\'e-invariant:
\[
\frac{\de}{\de t}\oint_{\zeta_t (\bX+\bX_\bU)\,}\boldsymbol{p}\cdot\de\bq=0
\,,
\]
where the curve $\zeta_t$ now moves along the \emph{total} phase-space vector field $\bX+\bX_\bU$.

\bigskip

\noindent\textbf{Ertel's theorem.}\quad By proceeding as in the corresponding treatment for the current-coupling scheme, one recognizes that the force
\begin{equation}\label{KNForce-PCS1}
\boldsymbol\Psi=\left(\mu_0^{-1}\nabla\times
\bB-q_h\bK\right) \times\bB+q_h(\nabla\cdot\bK)\bA
\end{equation}
provides the opportunity to generalize Ertel's relation for MHD \cite{Hide} to apply to the first pressure-coupling scheme. Indeed, upon following similar steps as those in Section \ref{Ertel-CCS}, one finds that the incompressible form of equation \eqref{KM-momap} yields the relation \eqref{Ertel}, {\color{black}with $\bar{\boldsymbol\omega}=\nabla\times\left(\bU-\rho^{-1}\!\int
f\,\bp\,\de^3\bv\right)$ and $\boldsymbol\Psi$ as given in \eqref{KNForce-PCS1}. Again, we kept the density $\rho$ in the expression of $\bar{\boldsymbol\omega}$ in order to provide correct dimensions; incompressible flows are always accompanied by $\rho=1$.}

\bigskip

\noindent\textbf{Cross helicities.}\quad
Notice that, upon denoting $\boldsymbol{\mathcal{W}}=\bU-\rho^{-1\!}\int\!f\,\boldsymbol{p}\,\de^3\bv$, both of the cross helicities
\[
\Lambda_1=\int\bU\cdot\bB\,\de^3\bq\,\qquad\Lambda_3=\int\boldsymbol{\mathcal{W}}\cdot\bB\,\de^3\bq
\]
possess nontrivial dynamics. Indeed, their equations of motion read as
\[
\frac{\de\Lambda_1}{\de t}=-m_h\int\!\rho^{-1}\left(\nabla\cdot\Bbb{P}\right)\cdot\bB\,\de^3\bq\,,\qquad
\frac{\de\Lambda_3}{\de t}=q_{h\!}\int\!\rho^{-1}\left(\bA\cdot\bB\right)\nabla\cdot\bK\,\de^3\bq\,.
\]
On the other hand, the following cross helicity is conserved:
\begin{equation}\label{CH-PCS1}
\Lambda_2=\int\boldsymbol{\Upsilon}\cdot\bB\,\de^3\bq\,,
\end{equation}
where we have denoted {\color{black}$\boldsymbol{\Upsilon}=\bU-m_h\rho^{-1}\bK$}. Upon noticing that $D_t(\rho^{-1\,}n)=\rho^{-1\!}\left(\partial_t n+\nabla\cdot(n\bU)\right)$, the conservation of $\Lambda_2$ is readily seen by computing
\[
\frac{\de\Lambda_2}{\de t}=\frac{\de\Lambda_3}{\de t}+q_{h\,}\frac{\de}{\de t}\int\!\rho^{-1\,}n\bA\cdot\bB\,\de^3\bq=0\,,
\]
where one considers the equation $\partial_t n+\nabla\cdot(n\bU)=-\nabla\cdot\bK$ arising from the zeroth moment of the Vlasov equation \eqref{PCS-MHD2}.

\section{Second pressure-coupling hybrid MHD scheme}\label{pressure-coupling2}

\subsection{Formulation of the model}

As mentioned in the previous section, the pressure-coupling MHD
scheme is conventionally obtained under the assumption that the hot
plasma component is rarefied. Upon denoting by $n$ the particle density of the hot component and by $m_c$ the cold
particle mass, this assumption reads as $n\ll\rho/m_c$. Then, in
order to avoid divergences in the mean \emph{velocity}
$\bV=n^{-1}\int\!\bv\,f\,\de^3\bv$ of the hot component, a small hot particle
density $n$ requires the hot \emph{momentum} $\bK=\int\!\bv\,f\,\de^3\bv$ to also be small. Thus,
it is customary to replace the total momentum $\rho\bU+m_h\bK$ by
simply $\rho\bU$, i.e. the cold fluid momentum. While this operation
is often performed on the equations of motion \cite{Cheng,ParkEtAl},
our approach makes this replacement directly in the variational
principle, resulting in agreement with \cite{Tronci2010}. The advantage of modelling in the
Lagrangian of the Euler-Poincar\'e variational principle is that it always produces circulation theorems.  This is the content of the Kelvin-Noether theorem of \cite{HoMaRa}. 

The assumption of a rarefied hot component {\color{black}may also
require} that the mean kinetic energy $m_h/2\int\!
n\left|\bV\right|^2\de^3\bq$ of the hot component is subtracted
from the corresponding total kinetic energy $m_h/2\int\!
f\left|\bv\right|^2\de^3\bq\,\de^3\bv$. This operation yields the second pressure-coupling scheme \cite{Tronci2010}
\begin{align}\label{PCS2-MHD1}
&\rho\frac{\partial \bU}{\partial
t}+\rho(\bU\cdot\nabla)\bU=-\nabla\mathsf{p}-m_h\,\nabla\cdot\!\int\!\left(\bv-\frac{\bK}{n}\right)\left(\bv-\frac{\bK}{n}\right)f\,\de^3\bv-\frac1{\mu_0}\bB\times\nabla\times\bB
\\\label{PCS2-MHD2}
&\frac{\partial f}{\partial t}+\left(\bv+\bU-\frac{\bK}{n}\right)\cdot\frac{\partial f}{\partial \bq}+\left(\!a_h\!\!\left(\bv-\frac{\bK}{n}\right)\times\bB{\color{black}-\nabla\bU\cdot\bv+\left(\nabla\frac{\bK}{n}\right)\cdot\left(\bv-\frac{\bK}{n}\right)}\!\!\right)\cdot\frac{\partial f}{\partial \bv}=0
\\\label{PCS2-MHD3}
&\frac{\partial \rho}{\partial t}+\nabla\cdot(\rho\,\bU)=0
\,,\qquad\ \frac{\partial \bB}{\partial
t}=\nabla\times\left(\bU\times\bB\right) \,.
\end{align}
We remark that neglecting all $\bU$- and $n^{-1}\bK$-terms in the
kinetic equation \eqref{PCS2-MHD2} and replacing $n^{-1}\bK\times\bB$ by $\bU\times\bB$  produces the hybrid MHD model in
\cite{KimEtAl,TaBrKi} (although the general Vlasov equation is adopted
here, rather than a {\color{black}drift-kinetic equation). Upon denoting $\bar{\Bbb{P}}=\int\!(\bv-\langle\bv\rangle)^{\otimes\,2}f\, \de^3\bv$ and $\langle\bv\rangle=\bK/n$, the model in \cite{KimEtAl,TaBrKi} can be derived by assuming $n\,\partial_t\langle\bv\rangle+n\left(\langle\bv\rangle\cdot\nabla\right)\langle\bv\rangle=-\nabla\cdot\bar{\Bbb{P}}+a_h(\bK-n\bU)\times\bB\simeq0$ in the current-coupling scheme \eqref{cc-hybrid-momentum}-\eqref{cc-hybrid-end}, so that Lorentz forces in \eqref{cc-hybrid-momentum} are replaced by a relative pressure term.

 Notice that the static equilibria of the above equations \eqref{PCS2-MHD1}-\eqref{PCS2-MHD3} coincide with those of the hybrid model in \cite{KimEtAl,TaBrKi} (for hot particles undergoing Vlasov dynamics), provided the equilibrium Vlasov distribution (usually denoted by $f_0$) is isotropic in the velocity coordinate, i.e. $\bK_0=\int\!\bv\,f_0\,\de^3\bv=0$.}

\subsection{Euler-Poincar\'e reduction by symmetry}

In order to obtain the Euler-Poincar\'e formulation of the hybrid model \eqref{PCS2-MHD1}-\eqref{PCS2-MHD3} for the second pressure-coupling scheme \cite{Tronci2010}, the Lagrangian \eqref{PCS-Lagrangian}  is transformed into
\begin{multline}
l(\bU,\bV,\bX,\rho,\bA,f)=\frac1{2}\int\!
\rho\,{\left|\bU\right|^2}\,\de^3\bq -\int\!\rho\,
\mathcal{U}(\rho)\,\de^3\bq-\frac1{2\mu_0}\int
\left|\nabla\times\bA\right|^2\de^3\bq
\\
+\int \!f\left(\frac{m_h}2
\left|\bu\right|^2+\frac{m_h}2
\left|\bu-\bv\right|^2-\frac{m_h}2\left|\boldsymbol{V}\right|^2+q_h\left(\bu+\bV\right)\cdot\bA
\right)\dvol \,,
\label{PCS2-Lagrangian}
\end{multline}
where we notice that the mean velocity $\bV$ appears as a new dynamical variable. The term $q_h\int \!f\,\bV\cdot\bA\,\dvol$ has been inserted in order to match the correct Lorentz force on the hot component \cite{Tronci2010}. {\color{black}As we shall see, the two $\bV-$terms  in the above Lagrangian correspond to subtracting the contributions of the mean velocity $\langle\bv\rangle=n^{-1}\bK=-\bV$.} Upon following the same reasoning as in the previous section, we realize that the fluid $\bU$-transport exerted by the cold fluid component on the Vlasov distribution $f$ of the hot particles must imply a $\bU$-transport of the mean hot velocity $\bV$. More particularly, we interpret the above Lagrangian as a functional of the type
\[
l:\mathfrak{X}_1(\Bbb{R}^3)\,\circledS\,\big(\mathfrak{X}_2(\Bbb{R}^3)\,\circledS\,\mathfrak{X}(\Bbb{R}^6)\big)\times
\operatorname{Den}(\Bbb{R}^3)\times\Omega^1(\Bbb{R}^3)\times
\operatorname{Den}(\Bbb{R}^6)\to\Bbb{R}
\,,\]
where $\mathfrak{X}_1(\Bbb{R}^3)$ and $\mathfrak{X}_2(\Bbb{R}^3)$ are two copies of the \emph{same} Lie algebra $\mathfrak{X}(\Bbb{R}^3)$ of vector fields, although they are denoted differently because the second is assumed to act trivially on the space $\operatorname{Den}(\Bbb{R}^3)\times\Omega^1(\Bbb{R}^3)$, containing the cold fluid density $\rho$ as well as the magnetic potential $\bA$. The first ({\color{black}outer}) semidirect-product symbol corresponds to fluid $\bU$-transport of both the mean velocity $\bV\in\mathfrak{X}_2(\Bbb{R}^3)$ and the phase-space vector field $\bX\in\mathfrak{X}(\Bbb{R}^6)$. On the other hand, the second ({\color{black}inner}) semidirect-product symbol corresponds to the $\bV$-transport exerted by the mean flow of the hot component on its corresponding phase-space velocity. At the group level, the unreduced Lagrangian is of the type
\begin{equation}\label{SDP-pc2}
L_{\rho_0,\bA_0,f_0}:T\!\left(\operatorname{Diff}_1(\Bbb{R}^3)\,\circledS\,\big(\!\operatorname{Diff}_2(\Bbb{R}^3)\,\circledS\operatorname{Diff}(\Bbb{R}^6)\big)\right)\to\Bbb{R}\,,
\end{equation}
where $(\rho_0,\bA_0,f_0)$ are the advected parameters.  Notice that similar arguments to those in Remark \ref{ConjAct} also apply here about the group actions involved in nested semidirect-product Lie group structures of this kind. 
{\color{black}
The first instance of nested semidirect-product Lie-group structures also occurred in plasma physics: in the discovery of the Lie-Poisson brackets dual to nested semidirect-product Lie algebras in models of Alfv\'en wave turbulence \cite {HaHoMo1985,Ho1985}. This construction was also used for hybrid Vlasov-fluid models in \cite{Tronci2010}. 
}
Further details can be found in \cite{GBHoPuRa}, where similar Lie group structures were shown to arise in polymer dynamics.

At this point, general geometric mechanics arguments ensure that the Euler-Poincar\'e variational principle $\delta\int_{t_0}^{t_1}l(\bU,\bV,\bX,\rho,\bA,f)\,\de t=0$ produces the following equations of motion:
\begin{align}\label{EP-PCS-2}
&\frac{\partial}{\partial t}\frac{\delta l}{\delta
{\bU}}+\pounds_{\bU}\,\frac{\delta l}{\delta \bU}=-\pounds_\bV\frac{\delta
l}{\delta \bV}+\frac{\delta
l}{\delta \bX}\star\bX+\left(\frac{\delta l}{\delta
\rho},\frac{\delta l}{\delta \bA},\frac{\delta l}{\delta
f}\right)\diamond_1\left(\rho,\bA,f\right)
\\\label{EP-PCS3-2}
&\frac{\partial \rho}{\partial t}+\pounds_{\bU}\, \rho=\,0\,, \quad\
\frac{\partial \bA}{\partial t}+\pounds_\bU\, \bA=0
\\\label{EP-PCS5}
&\frac{\partial}{\partial t}\frac{\delta l}{\delta
{\bV}}+\pounds_{\bV+\bU}\,\frac{\delta l}{\delta \bV}=\frac{\delta
l}{\delta \bX}\star\bX+\frac{\delta l}{\delta
f}\diamond_1f
\\\label{EP-PCS2-2}
&\frac{\partial}{\partial t}\frac{\delta l}{\delta
\bX}+\pounds_{\bX+\bX_{\bV+\bU}}\,\frac{\delta l}{\delta \bX}=
f\,\nabla_{(\bq,\bv)}\frac{\delta l}{\delta f}
\\\label{EP-PCS4-2}
&\frac{\partial f}{\partial t}+\pounds_{\bX+\bX_{\bV+\bU}}\, f=0
\end{align}

In order to see, how  equations \eqref{EP-PCS-2}-\eqref{EP-PCS4-2}  recover the second pressure coupling scheme \eqref{PCS2-MHD1}-\eqref{PCS2-MHD3}, it suffices to {\color{black} substitute}  the Lagrangian \eqref{PCS2-Lagrangian}. After computing
\[
\frac{\delta l}{\delta \bu}=m_h f\left(2\bu-\bv+a_h\,\bA\right),
\quad \frac{\delta l}{\delta \balpha}=0\,, \quad \frac{\delta
l}{\delta
f}=\frac{m_h}2|\bu|^2+\frac{m_h}2|\bu-\bv|^2-\frac{m_h}2\left|\bV\right|^2+q_h\left(\bu+\bV\right)\cdot\bA\,,
\]
the second component of equation \eqref{EP-PCS4} yields $\bu=\bv$ so that $\bX=(\bv,\balpha(\bq,\bv))$, similarly to the results in the previous section. Moreover, the first component of \eqref{EP-PCS4} reads as
\[
\partial_{t\,}\bp+\big(\left(\bv+\bV+\bU\right)\cdot\partial_\bq\big)\bp+(\ba\cdot\partial_\bv)\bp+\partial_{\bq\!}\left(\bv+\bV+\bU\right)
\cdot\bp=\partial_\bq\!\left(q_h(\bv+\bV)\cdot\bA-\frac{m_h}2|\bV|^2\right)\,.
\]
where we have denoted by $\ba=\balpha+\left(\bX_{\bV+\bU}\right)_2=\balpha+(\bv\cdot\nabla)(\bV+\bU)$ the total acceleration of the hot particles. After using standard vector identities, this equation yields the expression for the total force on the hot component:
\begin{equation*}\color{black}
m_h\ba=q_h\left(\bv+\bV\right)\times\bB-m_h\nabla\bU\cdot\bv-m_h\nabla\bV\cdot\left(\bv+\bV\right)\,.
\end{equation*}
Then, the total vector field $\bX+\bX_{\bV+\bU}=\left(\bv+\bV+\bU,\ba\right)$ is divergence-less and the Vlasov equation \eqref{EP-PCS4} becomes
\begin{equation}\color{black}
\frac{\partial f}{\partial t}+\left(\bv+\bV+\bU\right)\cdot\frac{\partial f}{\partial \bq}+\Big(a_h\left(\bv+\bV\right)\times\bB-\nabla\bU\cdot\bv-\nabla\bV\cdot\left(\bv+\bV\right)\Big)\cdot\frac{\partial f}{\partial \bv}=0\,.
\label{Vlasov-2nd-PCS}
\end{equation}
At this point, we observe that the constraint $\bV=-\bK/n$ is preserved by the dynamics. This is a direct consequence of the following 
\begin{proposition}\label{Momap-prop-PCS2}
With the notation of Lemma \ref{lemma-momap}, the Euler-Poincar\'e equations  \eqref{EP-PCS-2}-\eqref{EP-PCS4-2} yield
\begin{equation}
\left(\frac{\partial}{\partial t}+\pounds_{\bV+\bU}\right)\left(\frac{\delta l}{\delta \bV}
-i^*\!\left(\frac{\delta l}{\delta \bX}\right)\right)=0
\end{equation}
\end{proposition}
The proof proceeds analogously to that of Proposition \ref{Momap-prop} (see also Lemma \ref{lemma-momap}). Then, upon considering the Lagrangian \eqref{PCS2-Lagrangian}, the Euler-Poincar\'e equations \eqref{EP-PCS-2}-\eqref{EP-PCS4-2} preserve the constraint
\[
-\bV\int\! f\,\de^3\bv=\int\! \bv\,f\,\de^3\bv\,,
\]
which allows one to recover the Vlasov equation \eqref{PCS2-MHD2} of the second pressure coupling scheme in \cite{Tronci2010}.
Analogously, one can show that equation \eqref{EP-PCS-2} recovers the hybrid equation of motion of the same pressure coupling scheme, {\color{black}that is \eqref{PCS2-MHD1}}.
In order to show this, it suffices to verify that
\[
-\pounds_\bV\frac{\delta
l}{\delta \bV}+\frac{\delta l}{\delta
f}\diamond_1 f=-m_h\nabla\cdot\!\int\!\left(\bv-\frac{\bK}{n}\right)\left(\bv-\frac{\bK}{n}\right)f\,\de^3\bv=:-m_h\nabla\cdot\overline{\Bbb{P}}\,.
\]
This formula requires a lengthy but straightforward calculation that uses $\bV=-n^{-1}\bK$ and the well known relation
\[
\nabla\cdot\!\int\!\bv\bv\,f\,\de^3\bv=\nabla\cdot\left(n^{-1}\bK\bK+\int\!\left(\bv-n^{-1}\bK\right)\left(\bv-n^{-1}\bK\right)f\,\de^3\bv\right)
\]
between the absolute and relative pressure tensors.
Thus, in conclusion, we have proven the following.\medskip

\begin{theorem}
The hybrid pressure-coupling MHD scheme
 \eqref{PCS2-MHD1}-\eqref{PCS2-MHD3} arises from the
Euler-Poincar\'e variational principle
\[
\delta\int_{t_0}^{t_1}\! l(\bU,\bV,\bX,\rho,\bA,f)\,\de t=0
\]
with the reduced Lagrangian
\[
l:\mathfrak{X}_1(\Bbb{R}^3)\,\circledS\,\big(\mathfrak{X}_2(\Bbb{R}^3)\,\circledS\,\mathfrak{X}(\Bbb{R}^6)\big)\times
C^\infty(\Bbb{R}^3)^*\times\Omega^1(\Bbb{R}^3)\times
C^\infty(\Bbb{R}^6)^*\to\Bbb{R}
\]
as in \eqref{PCS2-Lagrangian} and variations
\begin{align*}
&\delta\!\left(\bU,\bV,\bX\right)=\partial_t\!\left(\mathbf{W},\bP,\mathbf{Z}\right)-\left(\pounds_{\bU}\mathbf{W},\pounds_{\mathbf{W}}\bV-\pounds_\bU\bP+\pounds_\bV\bP,
\pounds_{\bX_\mathbf{P+W}}\bX
-
 \pounds_{\bX_{\bV+\bU}}\mathbf{Z}
+
\pounds_{\bX}\mathbf{Z} \right)
\\
& \delta f=-\pounds_{\mathbf{Z}+\bX_\mathbf{P+W}}f
\,,\,\qquad
\delta\!\left(\rho,
\bA\right)=-\pounds_{\mathbf{W}}\left(\rho,\bA\right)
,
\end{align*}
in which the vector fields $\mathbf{P,W}\in\mathfrak{X}(\Bbb{R}^3)$ and
$\mathbf{Z}\in\mathfrak{X}(\Bbb{R}^6)$ vanish at the endpoints. This
variational principle is equivalent to the Euler-Poincar\'e
equations \eqref{EP-PCS-2}-\eqref{EP-PCS4-2} which hold for an
arbitrary hybrid Lagrangian.
\end{theorem}

\subsection{Discussion}
It is relevant to notice that equations \eqref{EP-PCS-2} and \eqref{EP-PCS2-2} yield  the following relation:
\begin{multline}\label{KM-momap-PCS2}
\left(\frac{\partial}{\partial t}+\pounds_{\bU}\right)\left(\frac{\delta l}{\delta \bU}-i^*\!\left(\frac{\delta l}{\delta \bX}\right)\right)=-\pounds_{\bV\!}\left(\frac{\delta l}{\delta \bV}-i^*\!\left(\frac{\delta l}{\delta \bX}\right)\right)+\rho\,\nabla\frac{\delta l}{\delta
\rho}
\\
-\frac{\delta l}{\delta
\bA}\times\nabla\times\bA+\left(\nabla\cdot\frac{\delta
l}{\delta \bA}\right)\bA\,.
\end{multline}
where one has used Lemma \ref{lemma-momap}. Upon inserting the Lagrangian \eqref{PCS2-Lagrangian}, the Lie derivatives in the right hand side cancel due to the constraint $\bV=-\bK/n$.

\bigskip

\noindent\textbf{Kelvin circulation laws.}\quad
The Kelvin-Noether conservation laws also hold for the equations \eqref{PCS2-MHD1}-\eqref{PCS2-MHD3}. Indeed, upon repeating the same steps as in the proof of Corollary \ref{KN-PCS1}, one finds the circulation laws
\begin{align}\label{PCS2-KN1}
&\frac{\de}{\de t}\oint_{\gamma_t(\bU)}\bU\cdot\de\bq
=-\oint_{\gamma_t(\bU)}\frac1\rho
\left(\frac1{\mu_0} \bB\times \nabla\times \bB
+
m_h\nabla\cdot\overline{\Bbb{P}}\right)\cdot\de\bq
\\ \label{PCS2-KN2}
&\frac{\de}{\de t}\oint_{\gamma_t(\bU)}\left(\bU-\frac1\rho\int
f\,\bp\,\de^3\bv\right)\cdot\de\bq
=
-\mu_0^{-1}\oint_{\gamma_t(\bU)}\frac1\rho\ 
\bB\times\nabla\times\bB
\cdot\de\bq \,,
\end{align}
where the second is a direct consequence of equation \eqref{KM-momap-PCS2}.
These results coincide with those found in \cite{Tronci2010} within the Lie-Poisson Hamiltonian setting. {\color{black} Now  taking the difference of the above two relations yields
\[
\frac{\de}{\de t}\oint_{\gamma_t(\bU)}\left(\frac1\rho\int
f\,\bp\ \de^3\bv\right)\cdot\de\bq
=
m_h\,\frac{\de}{\de t}\oint_{\gamma_t(\bU)}\frac\bK\rho\cdot\de\bq
=
-m_h\oint_{\gamma_t(\bU)}\frac1\rho
\left(\nabla\cdot\overline{\Bbb{P}}\right)\cdot\de\bq
\,,
\]
where we have used the zero-th moment equation $\partial_t n+\nabla\cdot(n\bU)=0$ associated to \eqref{Vlasov-2nd-PCS}. Indeed, together with equations \eqref{EP-PCS3-2}, this yields the following conserved circulation:
\[
\frac{\de}{\de t}\oint_{\gamma_t(\bU)}\!\left(1+\frac{n}\rho\right)\,\bA\cdot\de\bq=0\,.
\]
}
Moreover, the above circulation laws are accompanied by the following Poincar\'e invariant relation:
\[
\frac{\de}{\de t}\oint_{\zeta_t (\bX+\bX_{\bU-\bK/n})\,}\boldsymbol{p}\cdot\de\bq=0
\,,
\]
where the curve $\zeta_t$ moves along the total phase-space vector field $\bX+\bX_{\bU-\bK/n}$.

\bigskip


\noindent\textbf{Ertel's relation.}\quad
Proceeding as in Section \ref{Ertel-CCS}, {\color{black} taking the curl of the incompressible version of equation \eqref{KM-momap-PCS2} for the Lagrangian \eqref{PCS-Lagrangian}, or equivalently applying the Stokes theorem in the Kelvin-Noether relation \eqref{PCS2-KN2}  produces an Ertel relation of the form}
\[
D_t\left(\bar{\boldsymbol{\omega}}\cdot\nabla\alpha\right)-\left(\bar{\boldsymbol{\omega}}\cdot\nabla\right)D_t\alpha=-\mu_0^{-1\,}\nabla\alpha\cdot\nabla\times
\Big(\bB\times (\nabla\times\bB) \Big)
,\]
where $\alpha$ is an arbitrary scalar function and, {\color{black} upon keeping the mass density $\rho$ (equal to one for incompressible flows),}
{\color{black} 
\[
\bar{\boldsymbol{\omega}}=\nabla\times \left(\bU-m_h\frac\bK\rho\right)
.
\]
The Ertel relation above written in terms of $\bar{\boldsymbol{\omega}}$ has the same form as} the corresponding relation in \cite{Hide}, except that the vorticity $\bar{\boldsymbol{\omega}}$  {\color{black} here involves a velocity shift due to the mean specific momentum carried by the particles}.

\bigskip

\noindent\textbf{Cross helicities.}\quad
Upon denoting $\boldsymbol{\mathcal{W}}=\bU-\rho^{-1\!}\int\!f\,\boldsymbol{p}\,\de^3\bv$, it easy to see that the two cross helicities
\[
\Lambda_1=\int\bU\cdot\bB\,\de^3\bq
\quad\hbox{and}\quad
\Lambda_3=\int\boldsymbol{\mathcal{W}}\cdot\bB\,\de^3\bq
\]
possess the following dynamics
\[
\frac{\de\Lambda_1}{\de t}=-m_h\int\!\rho^{-1}\left(\nabla\cdot\overline{\Bbb{P}}\right)\cdot\bB\,\de^3\bq\,,\qquad
\frac{\de\Lambda_3}{\de t}=0\,,
\]
so that $\Lambda_3$ is now conserved by the hybrid dynamics of equations  \eqref{PCS2-MHD1}-\eqref{PCS2-MHD3}. 
{\color{black} Also, notice the conservation of the following modified magnetic helicity
\[
\mathcal{H}=\int\!\frac{n}\rho\,\mathbf{A}\cdot\bB\,\de^3\bq
\,,
\]
which allows to write the cross helicity invariant $\Lambda_3$ as
\[
\Lambda_3=\int\!\left(\bU-m_h\frac\bK\rho\right)\cdot\bB\,\de^3\bq
\]
}
The conservation law for $\color{black}\mathcal{H}$ provides an interesting opportunity to study the stability properties of this hybrid scheme. {\color{black} In particular, $\mathcal{H}$ does \emph{not} vanish for static equilibria. This means the energy-Casimir method may be applied for hybrid fluid equilibria that are analogous to the Chandrasekhar flows of inviscid MHD, \cite{HoMaRaWe1985}. }

\section{Summary and conclusions} \label{conclusions}

{\color{black} This paper has derived three different hybrid Vlasov-fluid plasma models by using the Euler-Poincar\'e approach first developed for the Maxwell-Vlasov plasma \cite{CeHoHoMa}.} After presenting the Euler-Poincar\'e approach for the Vlasov-multifluid plasma system, the discussion focused on three different schemes for deriving hybrid Vlasov-fluid MHD models. These comprised the current-coupling scheme and two pressure-coupling schemes. The first hybrid model was written on the direct product of two different diffeomorphism groups, as explained in Remark \ref{EPcc-DP}. The second one involved the more sophisticated construction of the semidirect-product diffeomorphism group discussed in Remark \ref{ConjAct}. Finally, a compound semidirect-product structure arose for the third hybrid model, see equation \eqref{SDP-pc2}.  In all three theories, Kelvin circulation theorems were presented for both the fluid motion and the hot particle dynamics on phase space, 
{\color{black} and the invariant cross-helicities were identified.}
A Legendre transform in each case would recover the Lie-Poisson results found \cite{Tronci2010}. Shifting to the {\color{black}drift-kinetic (or even gyrokinetic)} approximation would require another Lagrangian, which may also be derived systematically from the Lagrangian for Vlasov-MHD. 
{\color{black}
Summaries of the properties found here for the three different hybrid Vlasov-fluid plasma models are given below.

\medskip

\begin{framed}
\noindent
{\bfi Current-coupling MHD scheme}

\begin{itemize}
\item Equations of motion \eqref{cc-hybrid-momentum}-\eqref{cc-hybrid-end}:
\begin{align*}
&\rho\frac{\partial\bU}{\partial
t}+\rho\left(\bU\cdot\nabla\right)\bU = \left(q_h\,n\, \bU-\,q_h\,\bK+\frac1{\mu_0}\nabla\times\bB\right)\times\bB
-\nabla\mathsf{p}
\\
& \frac{\partial f}{\partial t}+\bv\cdot\frac{\partial f}{\partial
\bq}+a_h\left(\bv-\bU\right)\times\bB\cdot\frac{\partial f}{\partial
\bv}=0 
\\
& \frac{\partial\rho}{\partial
t}+\nabla\cdot\left(\rho\bU\right)=0
\,,\qquad
 \frac{\partial\bB}{\partial
t}=\nabla\times\left(\bU\times\bB\right)  \,;
\end{align*}
\item Kelvin circulation laws derived from  \eqref{Aeqn-CCS}-\eqref{KN-CCS}:
\begin{align*}
&\frac{\de}{\de t}\oint_{\gamma_t}\bU\cdot\de\bq=\oint_{\gamma_t}\frac1\rho
\left(q_h\,n\, \bU-\,q_h\,\bK+\frac1{\mu_0}\nabla\times\bB\right)\times\bB
\cdot\de\bq
\,,
\\
&
\frac{\de}{\de t}\oint_{\gamma_t}\left(1+\frac{n}{\rho}\right)\bA\cdot\de\bq=\oint_{\gamma_t}\frac1\rho
\Big(\nabla\cdot\left(n\,\bU-\bK\right)\!\Big)\bA\cdot\de\bq
\end{align*}
where $\gamma_t$ is any closed loop that moves with the fluid velocity $\bU$.

\pagebreak 

\item Magnetic and cross helicity invariants:
\[
\mathcal{H}=\int\bA\cdot\bB\,\de^3\bq
\,,\qquad
\Lambda=\int\bU\cdot\bB\,\de^3\bq
\]

\item Approximation: same as ideal MHD; this yields the Lagrangian \eqref{CCS-Lagrangian}
\end{itemize}
\end{framed}

\medskip

\begin{framed}\noindent
{\bfi Pressure-coupling MHD scheme -- first variant}

\begin{itemize}
\item Equations of motion  \eqref{PCS-MHD1}-\eqref{PCS-MHD3}:
\begin{align*}
&\rho\frac{\partial \bU}{\partial
t}+\rho(\bU\cdot\nabla)\bU=-\nabla{\sf p}-m_h\nabla\cdot\Bbb{P} -\frac1{\mu_0}\bB\times\nabla\times \bB
\\
&\frac{\partial f}{\partial
t}+\left(\boldsymbol{U}+\bv\right)\cdot\frac{\partial f}{\partial
\bq}-\frac{\partial f}{\partial
\bv}\cdot\nabla\boldsymbol{U}\cdot\bv+a_h\,{\bv}\times
\bB\cdot\frac{\partial f}{\partial\bv}=0
\\
&\frac{\partial \rho}{\partial t}+\nabla\cdot(\rho\,\bU)=0
\,,\qquad \frac{\partial \bB}{\partial
t}=\nabla\times\left(\bU\times\bB\right) \,.
\end{align*}

\item Kelvin circulation laws \eqref{KN1}-\eqref{KN2}:
\begin{align*}
&\frac{\de}{\de
t}\oint_{\gamma_t}\bU\cdot\de\bq=-\oint_{\gamma_t}\frac1\rho
\left(\frac1{\mu_0}\bB\times\nabla\times
\bB+m_h\nabla\cdot\Bbb{P}\right)\cdot\de\bq
\\
&\frac{\de}{\de t}\oint_{\gamma_t}\frac\bK\rho\cdot\de\bq=\oint_{\gamma_t}\frac1\rho
\Big(a_h\bK\times\bB-\nabla\cdot\Bbb{P}\Big)\cdot\de\bq 
\\
&
\frac{\de}{\de t}\oint_{\gamma_t}\left(1+\frac{n}{\rho}\right)\bA\cdot\de\bq=-\oint_{\gamma_t}\frac1\rho
\left(\nabla\cdot\bK\right)\bA\cdot\de\bq
\,;
\end{align*}

\item Magnetic and cross helicity invariants:
\begin{equation*}
\mathcal{H}=\int\bA\cdot\bB\,\de^3\bq
\,,\qquad
\Lambda=\int\!\left(\bU-m_h\frac\bK\rho\right)\cdot\bB\,\de^3\bq
\end{equation*}

\item Approximation: neglects minimal coupling term $\int\! n\,\bU\cdot\bA\,\de^3\bq$ in the Lagrangian \eqref{CCS-Lagrangian}.
\end{itemize}
 \end{framed}
 
\medskip
 
\begin{framed}\noindent
{\bfi Pressure-coupling MHD scheme -- second variant} 

\medskip
\begin{itemize}
\item Equations of motion  \eqref{PCS2-MHD1}-\eqref{PCS2-MHD3}:
 \begin{align*}
&\rho\frac{\partial \bU}{\partial
t}+\rho(\bU\cdot\nabla)\bU=-\nabla\mathsf{p}-m_h\,\nabla\cdot\bar{\Bbb{P}}-\frac1{\mu_0}\bB\times\nabla\times\bB
\\
&\frac{\partial f}{\partial t}+\left(\bv+\bU-\frac{\bK}{n}\right)\cdot\frac{\partial f}{\partial \bq}+\left(\!a_h\!\!\left(\bv-\frac{\bK}{n}\right)\times\bB-\nabla\bU\cdot\bv+\left(\nabla\frac{\bK}{n}\right)\cdot\left(\bv-\frac{\bK}{n}\right)\!\!\right)\cdot\frac{\partial f}{\partial \bv}=0
\\
&\frac{\partial \rho}{\partial t}+\nabla\cdot(\rho\,\bU)=0
\,,\qquad\ \frac{\partial \bB}{\partial
t}=\nabla\times\left(\bU\times\bB\right) \,.
\end{align*}
\item Kelvin circulation laws \eqref{PCS2-KN1}-\eqref{PCS2-KN2}:
\begin{align*}
&\frac{\de}{\de
t}\oint_{\gamma_t}\bU\cdot\de\bq=-\oint_{\gamma_t}\frac1\rho
\left(\frac1{\mu_0}\bB\times\nabla\times
\bB+m_h\nabla\cdot\bar{\Bbb{P}}\right)\cdot\de\bq
\\
&\frac{\de}{\de t}\oint_{\gamma_t}\frac\bK\rho\cdot\de\bq
=
-\oint_{\gamma_t}\frac1\rho
\left(\nabla\cdot\overline{\Bbb{P}}\right)\cdot\de\bq
\,,\quad\
\frac{\de}{\de t}\oint_{\gamma_t}\left(1+\frac{n}{\rho}\right)\bA\cdot\de\bq=0
\end{align*}

\item Magnetic and cross helicity invariants:
\begin{equation*}
\mathcal{H}=\int\!\left(1+\frac{n}\rho\right)\bA\cdot\bB\,\de^3\bq
\,,\qquad
\Lambda=\int\!\left(\bU-m_h\frac\bK\rho\right)\cdot\bB\,\de^3\bq
\end{equation*}

\item  Approximation: neglects $\int\! n\,\bU\cdot\bA\,\de^3\bq$ as well as  mean flow terms in the Lagrangian \eqref{CCS-Lagrangian}
\end{itemize}
\end{framed}
}

The Euler-Poincar\'e approach provided the means of comparing the geometrical properties of these three hybrid Vlasov-fluid plasma schemes in the same framework. This framework allowed the identification and comparison of the geometric relationships within each scheme that were shared by the others.  We expect that this framework will be useful in other modelling contexts. For example, one may imagine using the Euler-Poincar\'e framework, (i) in the comparison and selection of Vlasov-fluid hybrid models, (ii) in the validation of previous derivations, (iii) in making choices among the schemes in various physical regimes, and (iv) as a basis for performing other derivations obtained by modelling in the Lagrangian. 
From the physical viewpoint the roles of heat exchange and other advected quantities should also lead to interesting effects in future investigations.
For example, the introduction of another advected quantity would produce an explicit Ertel theorem for the evolution of potential vorticity. Of course, Ertel's theorem is an immediate result of the Euler-Poincar\'e theory for any hybrid fluid-Vlasov model with advected quantities. However, it was discussed here only in the simple case of incompressible hybrid fluid flows.

In another direction for future research in the context of potential vorticity, one may use Ertel's theorem to investigate the evolution of the gradient of the potential vorticity, as studied recently in geophysical fluid dynamics in terms of the vector $\boldsymbol{\mathcal{B}}=\nabla Q(q)\times\nabla\alpha$, \cite{HoGi}. Here $Q$ is an arbitrary function and $q=\bar{\boldsymbol{\omega}}\cdot\nabla\alpha$ is the potential vorticity arising in a given fluid model. Upon considering an advected function $\alpha$ such that $\partial_t\alpha+\bU\cdot\nabla\alpha=0$, Ertel's theorem was shown in \cite{HoGi} to produce the dynamics of the vector $\boldsymbol{\mathcal{B}}$ in the form
\[
\frac{\partial\boldsymbol{\mathcal{B}}}{\partial t}-\nabla\times\left(\boldsymbol{\mathcal{U}}\times\boldsymbol{\mathcal{B}}\right)=\nabla\times\boldsymbol{\Phi}
\,,\]
where $\boldsymbol{\mathcal{U}}=\bU-q^{-1}\big(\alpha\nabla\times\boldsymbol{\Psi}
\big)$ and 
$
\nabla\times\boldsymbol{\Phi}=\nabla\alpha\times\nabla\left(q\,Q'\,\nabla\cdot\boldsymbol{\mathcal{U}}\right).
$
Evidently, the divergenceless vector $\boldsymbol{\mathcal{D}}=\nabla\times\boldsymbol{\Phi}$ breaks the frozen-in condition of the vector $\boldsymbol{\mathcal{B}}$ and it thus affects the stretching properties that are governed by the left-hand side of the equation for the vector $\boldsymbol{\mathcal{B}}$, see \cite{HoGi}. The  identification of the vorticity $\bar{\boldsymbol{\omega}}$ and the force $\boldsymbol{\Psi}$ in each of the hybrid models discussed here would provide interesting opportunities to study the dynamics of the gradients of the potential vorticity. This would perhaps lead to the production of fronts and other fine structures at high wave numbers that develop by stretching of the $\boldsymbol{\mathcal{B}}$-vector.

\medskip

\paragraph{Acknowledgements.}
 We are grateful to C. David Levermore for many inspiring discussions on these and related topics over the years. This paper is dedicated to him on the occasion of his sixtieth birthday. Happy birthday, Dave! DDH was partially supported by the Royal Society of London Wolfson Scheme and the European Research Council Advanced Investigator Grant.

\medskip

\appendix

\section{Proof of Lemma \ref{lemma-momap}}\label{Appendix}
The map $i^*$ is easily seen to be a momentum map arising from the dual of the Lie algebra inclusion $i:\bU\mapsto\bX_\bU$. 
Upon denoting $\boldsymbol\Xi=\delta l/\delta\bX$, the momentum map property
\[
\big\{F, \left\langle i^*\!\left(\boldsymbol\Xi\right),\bU\right\rangle\!\big\}_{\mathfrak{X}^*(\Bbb{R}^6)\!}=\bU_{\!\mathcal{F}(\mathfrak{X}^*(\Bbb{R}^6))}\!\left[F\right]
\]
can be verified explicitly. Here, $\bU_{\!\mathcal{F}(\mathfrak{X}^*(\Bbb{R}^6))}\left[F\right]$ denotes evaluation on the functional $F(\boldsymbol\Xi)$ of the infinitesimal generator $\bU_{\!\mathcal{F}(\mathfrak{X}^*(\Bbb{R}^6))}$ of the $\operatorname{Diff}(\Bbb{R}^3)-$action on the space of functionals $\mathcal{F}(\mathfrak{X}^*(\Bbb{R}^6))$ on the one-form densities in $\mathfrak{X}^*(\Bbb{R}^6)$. Upon using the right Lie-Poisson bracket on $\mathfrak{X}^*(\Bbb{R}^6)$, one computes
\begin{align*}
\big\{F, \left\langle i^*\!\left(\boldsymbol\Xi\right),\bU\right\rangle\!\big\}_{\mathfrak{X}^*(\Bbb{R}^6)\!}=&\left\langle\boldsymbol\Xi,\left[\frac{\delta F}{\delta \boldsymbol\Xi},\frac{\delta}{\delta\boldsymbol\Xi}\left\langle i^*\!\left(\boldsymbol\Xi\right),\bU\right\rangle\right]\right\rangle_{\!\mathfrak{X}^*(\Bbb{R}^6)\!}
\\
=&\left\langle\boldsymbol\Xi,\left[\frac{\delta F}{\delta \boldsymbol\Xi},\frac{\delta}{\delta\boldsymbol\Xi}\left\langle \boldsymbol\Xi,i(\bU)\right\rangle\right]\right\rangle_{\!\mathfrak{X}^*(\Bbb{R}^6)\!}
\\
=&\left\langle\boldsymbol\Xi,\,\pounds_{i(\bU)\,}\frac{\delta F}{\delta \boldsymbol\Xi}\right\rangle_{\!\mathfrak{X}^*(\Bbb{R}^6)\!}
\\
=&-\left\langle\!\pounds_{\bX_\bU}\boldsymbol\Xi,\,\frac{\delta F}{\delta \boldsymbol\Xi}\right\rangle_{\!\mathfrak{X}^*(\Bbb{R}^6)\!}
\\
=&\ 
\bU_{\!\mathcal{F}(\mathfrak{X}^*(\Bbb{R}^6))}\!\left[F\right].
\end{align*}
where $\left[\cdot,\cdot\right]$ denotes minus the Jacobi-Lie bracket on $\mathfrak{X}(\Bbb{R}^6)$.

The rest of the proof proceeds in two steps. First, 
\begin{align*}
\int\!\left(\pounds_{\bX_{\bU}}\frac{\delta l}{\delta \bX}\right)_{\!1}\de^3\bv=&
\int\!\left(\bU\cdot\partial_\bq+(\bv\cdot\nabla)\bU\cdot\partial_\bv\right)\frac{\delta l}{\delta \bu}\,\de^3\bv
+\int\!\left(\nabla\bU\cdot\frac{\delta l}{\delta \bu}+\partial_\bq\!\left((\bv\cdot\nabla)\bU\right)\cdot\frac{\delta l}{\delta \balpha}
\right.
\\
&\left.+\Big(\nabla\cdot\bU+\partial_\bv\cdot\left((\bv\cdot\nabla)\bU\right)\Big)\frac{\delta l}{\delta \bu}\right)\de^3\bv
\\
=&
\int\!\left(\bU\cdot\partial_\bq\right)\frac{\delta l}{\delta \bu}\,\de^3\bv+
\int\!\left((\bv\cdot\nabla)\bU\cdot\partial_\bv\right)\frac{\delta l}{\delta \bu}\,\de^3\bv+\int\!\big((\bv\cdot\nabla)\nabla\bU\big)\cdot\frac{\delta l}{\delta \balpha}\,\de^3\bv
\\
&+\int\!\left(\nabla\bU\cdot\frac{\delta l}{\delta \bu}
+\left(\nabla\cdot\bU\right)\frac{\delta l}{\delta \bu}\right)\de^3\bv
-\int\!\left(
\left((\bv\cdot\nabla)\bU\cdot\partial_\bv\right)\frac{\delta l}{\delta \bu}\right)\de^3\bv
\\
=&\
\pounds_{\bU}\int\!\frac{\delta l}{\delta \bu}\,\de^3\bv+\int\!\big((\bv\cdot\nabla)\nabla\bU\big)\cdot\frac{\delta l}{\delta \balpha}\,\de^3\bv\,.
\end{align*}
One also has
\begin{align*}
\int\!(\bv\cdot\partial_\bq)\left(\pounds_{\bX_{\bU}}\frac{\delta l}{\delta \bX}\right)_{\!2}\de^3\bv=&
\int\!(\bv\cdot\partial_\bq)\left(\!\big(\bU\cdot\partial_\bq+(\bv\cdot\nabla)\bU\cdot\partial_\bv\big)\frac{\delta l}{\delta \balpha}\right)\de^3\bv
\\
&+\int\!(\bv\cdot\partial_\bq)\left(\partial_\bv\big((\bv\cdot\nabla)\bU\big)\cdot\frac{\delta l}{\delta \balpha}\right)\de^3\bv
\\
&+\int\!(\bv\cdot\partial_\bq)\left(\!\Big(\nabla\cdot\bU+\partial_\bv\cdot\big((\bv\cdot\nabla)\bU\big)\Big)\frac{\delta l}{\delta \balpha}\right)\de^3\bv
\end{align*}
Then, for each term, one computes
\begin{align*}
&\int\!(\bv\cdot\partial_\bq)\left(\!\big(\bU\cdot\partial_\bq\big)\frac{\delta l}{\delta \balpha}\right)\de^3\bv=\bU\cdot\nabla\!\int\!(\bv\cdot\partial_\bq)\frac{\delta l}{\delta \balpha}\,\de^3\bv+\int\!\operatorname{Tr}\!\Big((\nabla\bU)(\bv\partial_\bq)\Big)\frac{\delta l}{\delta \balpha}\,\de^3\bv\,,
\\
&\int\!(\bv\cdot\partial_\bq)\left(\!\Big(\partial_\bv\big((\bv\cdot\nabla)\bU\big)\Big)\frac{\delta l}{\delta \balpha}\right)\,\de^3\bv=\nabla\bU\!\cdot\!\int\!(\bv\cdot\partial_\bq)\frac{\delta l}{\delta \balpha}\,\de^3\bv+\int\!\big((\bv\cdot\nabla)\nabla\bU\big)\cdot\frac{\delta l}{\delta \balpha}\,\de^3\bv
\\
&\int\!(\bv\cdot\partial_\bq)\left(\!\big(\nabla\cdot\bU\big)\frac{\delta l}{\delta \balpha}\right)\de^3\bv
=\int\!\Big((\bv\cdot\nabla)\big(\nabla\cdot\bU\big)\Big)\frac{\delta l}{\delta \balpha}\,\de^3\bv+(\nabla\cdot\bU)\int\!(\bv\cdot\partial_\bq)\frac{\delta l}{\delta \balpha}\,\de^3\bv
\end{align*}
and
\begin{align*}
\int&(\bv\cdot\partial_\bq)\left(\Big(\partial_\bv\cdot\big((\bv\cdot\nabla)\bU\big)\Big)\frac{\delta l}{\delta \balpha}\right)\de^3\bv=\partial_\bq\cdot\int\!\Big(\partial_\bv\cdot\big((\bv\cdot\nabla)\bU\big)\Big)\bv\frac{\delta l}{\delta \balpha}\,\de^3\bv
\\
=&
-\nabla\cdot\!\int\!\big((\bv\cdot\nabla)\bU\big)\frac{\delta l}{\delta \balpha}\,\de^3\bv-\int\!(\bv\cdot\partial_\bq)\left(\!\big((\bv\cdot\nabla)\bU\cdot\partial_\bv\big)\frac{\delta l}{\delta \balpha}\right)\de^3\bv
\\
=&-\int\!\big((\bv\cdot\nabla)\nabla\bU\big)\cdot\frac{\delta l}{\delta \balpha}\,\de^3\bv-\int\!\operatorname{Tr}\!\Big((\nabla\bU)(\bv\partial_\bq)\Big)\frac{\delta l}{\delta \balpha}\,\de^3\bv-\int\!(\bv\cdot\partial_\bq)\left(\!\big((\bv\cdot\nabla)\bU\cdot\partial_\bv\big)\frac{\delta l}{\delta \balpha}\right)\de^3\bv\,.
\end{align*}
Thus, in conclusion
\[
\int\!\left(\pounds_{\bX_{\bU}}\frac{\delta l}{\delta \bX}\right)_{\!1}\de^3\bv
-\int\!(\bv\cdot\partial_\bq)\left(\pounds_{\bX_{\bU}}\frac{\delta l}{\delta \bX}\right)_{\!2}\de^3\bv
-\pounds_{\bU}\int\!\frac{\delta l}{\delta \bu}\,\de^3\bv+\pounds_{\bU}\int\!(\bv\cdot\partial_\bq)\frac{\delta l}{\delta \balpha}\,\de^3\bv=0
\]
which completes the proof. $\blacksquare$

\end{document}